\documentclass[prl,twocolumn,preprintnumbers,amsmath,amssymb,floatfix]{revtex4}
\usepackage{graphicx}
\usepackage{graphics}
\usepackage{psfrag}
\usepackage{hyperref}
\usepackage{multirow}
\usepackage[sort&compress]{natbib}
\usepackage{amssymb}
\usepackage{amsmath}
\usepackage{amsthm}
\usepackage{bbold}
\usepackage{bbm}
\usepackage{enumerate}

\newcommand{\rmi}{{\rm i}}

\usepackage{xcolor}

\begin{document}

\hypersetup{pdftitle={Unconventional Superconductivity in Luttinger Semimetals: Theory of Complex Tensor Order and the Emergence of the Uniaxial Nematic State}}
\title{Unconventional Superconductivity in Luttinger Semimetals:\\ Theory of Complex Tensor Order and the Emergence of the Uniaxial Nematic State}

\author{Igor Boettcher}
\affiliation{Department of Physics, Simon Fraser University, Burnaby, British Columbia, Canada, V5A 1S6}
\author{Igor F. Herbut}
\affiliation{Department of Physics, Simon Fraser University, Burnaby,  British Columbia, Canada, V5A 1S6}

\begin{abstract}
We investigate unconventional superconductivity in three-dimensional electronic systems with the chemical potential close to a quadratic band touching point in the band dispersion. Short-range interactions can lead to d-wave superconductivity, described by a complex tensor order parameter. We elucidate the general structure of the corresponding Ginzburg--Landau free energy and apply these concepts to the case of an isotropic band touching point. For vanishing chemical potential, the ground state of the system is given by the superconductor analogue of the uniaxial nematic state, which features line nodes in the excitation spectrum of quasiparticles. In contrast to the theory of real tensor order in liquid crystals, however, the ground state is selected here by the sextic terms in the free energy. At finite chemical potential, the nematic state has an additional instability at weak coupling and low temperatures. In particular, the one-loop coefficients in the free energy indicate that at weak coupling genuinely complex orders, which break time-reversal symmetry, are energetically favored. We relate our analysis to recent measurements in the half-Heusler compound YPtBi and discuss the role of the cubic crystal symmetry.
\end{abstract}

\maketitle

Three-dimensional electronic systems in which spin-orbit coupling is strong enough to produce band inversion so that two
bands touch quadratically near or at the Fermi level are interesting for a number of reasons. Opening a gap in the spectrum by applying strain, for example, famously yields a topologically nontrivial insulating state \cite{RevModPhys.83.1057}. Coulomb interaction sets the dominant energy scale near the touching point, and the ground state has been argued to become a non-Fermi liquid or to break some of the spatial symmetries \cite{abrikosov,abrben,moon,PhysRevLett.113.106401,JanssenHerbut,ArmitageReview}. Several aspects of tensorial magnetism have been explored theoretically and experimentally in the class of pyrochlore iridates \cite{PhysRevLett.96.087204,MachidaNature,doi:10.1143/JPSJ.80.044708,PhysRevLett.106.217204,BalentsReview,savary,PhysRevB.91.115124,murray,PhysRevLett.117.056403,kondo,OngNature,PhysRevB.95.085120,PhysRevB.95.075149}. More recently, superconductivity has been added to the list of phenomena that attract attention, particularly with the non-centrosymmetric half-Heusler alloys promising a pathway to novel unconventional and topological superconductivity \cite{GOLL20081065,ChadovNatMater,PhysRevB.84.220504,PhysRevB.86.064515,PhysRevB.87.184504,Pan2013,BayLowT,Xu2014,Nakajima2015,PhysRevB.93.205138,PhysRevLett.116.137001,PhysRevLett.116.177001,2016arXiv160303375K,SmidmanReview,2017arXiv170101553K,PhysRevLett.118.127001}.
Since the electrons occupying the inverted bands have total angular momentum of 3/2, this allows for Cooper pairs with total (integer) spin ranging from zero to three.

Motivated in particular by the recent observation of the linear temperature dependence of the penetration depth in YPtBi \cite{2016arXiv160303375K}, in this Letter we address the following basic problem. Assuming the simplest, maximally symmetric single-particle Luttinger Hamiltonian \cite{luttinger}, and the most general symmetry-allowed contact interactions between such spin-3/2 electrons, what is the ensuing superconducting state? This idealization is actually not far from reality for YPtBi, where terms that break particle-hole, full rotation, and inversion symmetry are all of the order of ten percent and lower.

Besides the obvious possibility of s-wave superconductivity, the only other superconducting order parameter that is finite at the point of the quadratic band touching (QBT) is the ($l=2$) d-wave state, which is described by a complex order parameter $\phi$ which transforms as an irreducible second-rank tensor under rotations. The intriguing interplay between complex and tensorial character of the d-wave state dictates the nature of its phase structure. Here we discuss and derive the corresponding Ginzburg--Landau (GL) expansion of the free energy.
The theory displays a number of novel features which are absent in the formally related GL theory for nematic order in liquid crystals \cite{BookDeGennes}. In particular, the crucial cubic term, $\mbox{tr}(\phi ^3)$, which there favors the uniaxial nematic state, is forbidden here by the particle number $\text{U}(1)$ symmetry, and the transition is in turn governed by the quartic and sextic terms. We expound here that only a few of these terms are independent, implying a transparent form of the GL theory with clear physical consequences.

Our GL free energy shows that at weak coupling with the accompanying very low critical temperature ($T_{\rm c} \ll \mu$), the transition at the mean-field level is continuous, and into a particular \emph{complex} (time-reversal symmetry breaking) configuration. A similar, although not identical, conclusion was reached in Ref. \cite{PhysRevLett.116.177001}, where the search for the energetically best configuration was constrained by the cubic symmetry from the outset \cite{RevModPhys.63.239}. As the coupling constant is increased and $T_c$ raised towards $\mu$, the result changes in two crucial respects: 1) a particular quartic term changes sign and thus causes the preferred order parameter to become \emph{real}, 2) the superconducting transition itself becomes discontinuous. The most striking result is that in a large portion of the phase diagram the lowest energy is achieved by the uniaxial nematic state that breaks $\text{U}(1)$ and rotational symmetry while preserving time reversal. This particular superconducting state features line nodes in the excitation spectrum, and therefore, if extending to low temperatures, would indeed display the observed linear temperature dependence of the penetration depth.

\paragraph{Invariant theory for complex tensor.} We first lay out the general theory of complex tensor order in three dimensions, before turning to the particular realization in Luttinger semimetals. To this purpose consider a system with microscopic interactions featuring rotation symmetry and particle number conservation, manifested as a global $\text{SO}(3)\times \text{U}(1)$ symmetry. Assume further the existence of a complex order parameter $\phi_{ij}$ with $i,j=1,2,3$ that transforms as a symmetric irreducible second-rank tensor under rotations, i.e. $\phi_{ij}=\phi_{ji}$, $\delta_{ij}\phi_{ij}=0$. (Under $\vec{x}\mapsto R\vec{x}$, $\phi \to R\phi R^{\rm T}$, with $R\in\text{SO}(3)$.)
The global $\text{SO}(3)\times \text{U}(1)$ invariance of the theory then implies that two order parameters $\phi$ and $\phi'$ are physically equivalent if there exists an $R$ such that $R\phi R^{\rm T}=e^{\rmi \alpha}\phi'$ with $e^{\rmi \alpha}$ a phase factor.

The most general GL free energy $F(\phi)$ describing complex tensor order can be constructed with the help of the following fact from invariant theory \cite{SpencerRivlin,ARTIN1969532,PROCESI1976306,Matteis2008}: Let $f(\phi)$ be a polynomial function of the complex symmetric traceless matrix $\phi$ that is invariant under $\phi\to R \phi R^{\rm T}$. Then $f(\phi)$ is a polynomial in only \emph{eight} invariants:
\begin{align}
 \nonumber I_1 &=\mbox{tr}(\phi^\dagger\phi),\ I_2 = \mbox{tr}(\phi^2),\ I_3=\mbox{tr}(\phi^\dagger{}^2),\\
 \nonumber I_4 &=\mbox{tr}(\phi^3),\ I_5 = \mbox{tr}(\phi^\dagger{}^3),\ I_6=\mbox{tr}(\phi^2\phi^\dagger),\\
 \label{eq1} I_7&=\mbox{tr}(\phi^\dagger{}^2\phi),\ I_8 = \mbox{tr}(\phi^\dagger\phi\phi^\dagger\phi),
\end{align}
 which comprise the so-called integrity basis of $\text{SO}(3)$. Furthermore, only seven of these invariants are actually functionally independent; $I_8$, for example, can be expressed as a non-polynomial function of the other seven. Consider an expansion of $F(\phi)$ in powers of $\phi$ and denote by $\mathcal{O}_n$ the set of independent terms that appear to $n$th power in $\phi$.
Then $\text{SO}(3)\times \text{U}(1)$ symmetry dictates
\begin{align}
 \label{eq2} \mathcal{O}_2 &= \{ I_1\},\\
 \label{eq3} \mathcal{O}_4 &= \{ I_1^2,\ I_2I_3,\ I_8\},\\
\label{eq4} \mathcal{O}_6 &= \{ I_1^3,\ I_1I_2I_3,\ I_4I_5,\ I_6I_7,\ I_1I_8\}.
\end{align}
To octic order, twelve terms are allowed. We emphasize the remarkable reduction of independent terms in comparison to all sextic and octic terms that naively can be constructed from $\phi$ respecting  $\text{SO}(3)\times \text{U}(1)$, such as $\mbox{tr}(\phi^2\phi^\dagger{}^2\phi\phi^\dagger)$ and $\mbox{tr}(\phi^2\phi^\dagger{}\phi\phi^\dagger{}^2\phi\phi^\dagger)$, for instance.

Any symmetric traceless $3\times 3$ matrix $\phi$, on the other hand, can be written as
\begin{align}
 \label{eq5}  \phi_{ij} = \Delta_a M^a_{ij},
\end{align}
with five components $\Delta_a\in \mathbb{C}$ and the real Gell-Mann matrices \cite{JanssenHerbut} given by
\begin{align}
 \nonumber M^1 &= \begin{pmatrix} 1 & 0 & 0 \\ 0 & -1 & 0 \\ 0 & 0 & 0 \end{pmatrix},\ M^2 = \frac{1}{\sqrt{3}} \begin{pmatrix} -1 & 0 & 0 \\ 0 & -1 & 0 \\ 0 & 0 & 2 \end{pmatrix},\\
 \label{eq6} M^3 &= \begin{pmatrix} 0 & 0 & 1 \\ 0 & 0 & 0 \\ 1 & 0 & 0 \end{pmatrix},\ M^4 = \begin{pmatrix} 0 & 0 & 0 \\ 0 & 0 & 1 \\ 0 & 1 & 0  \end{pmatrix},\ M^5 = \begin{pmatrix} 0 & 1 & 0 \\ 1 & 0 & 0 \\ 0 & 0 & 0 \end{pmatrix}.
\end{align}
The parametrization in Eq. (\ref{eq5}) with $\vec{\Delta}=(\Delta_1,\Delta_2,\Delta_3,\Delta_4,\Delta_5)$ allows us to write
\begin{align}
 \label{eq7} I_1 = 2|\vec{\Delta}|^2,\ I_2I_3 = 4(\vec{\Delta}^2)(\vec{\Delta}^2)^*.
\end{align}
These two expressions are thus invariant under the larger class of $\text{SO}(5)\times \text{U}(1)$ transformations applied to the five-component object $\vec{\Delta}$. In particular, assume $F(\phi)$ is such that $I_2I_3$ needs to be maximized. Due to $|\vec{\Delta}^2|\leq |\vec{\Delta}|^2$ this implies that $\vec{\Delta}$ is actually real (up to an overall phase factor). The real symmetric matrix $\phi$ can then be rotated into its eigenframe and is fully described by two real eigenvalues. In contrast, if $F(\phi)$ is such that $I_2I_3$ is to be minimized, the optimal configuration will be \emph{genuinely complex}, and thus break time-reversal symmetry. Indeed, due to $|\vec{\Delta}^2|\geq 0$ the best choice is $\vec{\Delta}^2=\sum_a \Delta_a^2=0$. However, this being a sum of squares implies that the components $\Delta_a$ are necessarily complex numbers. In Tab. \ref{TabDelta} we list a few configurations and a selection of their invariants.

\renewcommand{\arraystretch}{1.6}
\begin{table}[t]
\begin{tabular}{|c||c|c|c|c|c|}
\hline $\vec{\Delta}$ & $|\vec{\Delta}|^2$ & $\vec{\Delta}^2$ & $\mathcal{Q}/\Delta^4$ &  $\mathcal{S}/\Delta^6$ & comment \\
 \hline\hline $\Delta(1,0,0,0,0)$ & $\Delta^2$ & $\Delta^2$ & $1$ & 0 & \\
\hline $\Delta(0,1,0,0,0)$ & $\Delta^2$ & $\Delta^2$ & $1$ & $40/3$ & $\begin{array}{c} \text{uniaxial} \\  \text{nematic}\end{array}$\\
 \hline $\frac{\Delta}{\sqrt{2}}(1,1,0,0,0)$ & $\Delta^2$ & $\Delta^2$ & $1$ & $20/3$ & $d_1+d_2$ \\
 \hline\hline $\frac{\Delta}{\sqrt{2}} (1,\rmi,0,0,0)$ & $\Delta^2$ & 0 & $2/3$& $8/3$ & $d_1+\rmi d_2$\\
 \hline $\frac{\Delta}{\sqrt{2}} (0,0,1,\rmi,0)$ & $\Delta^2$ & 0 & $1$ & $0$ & $d_3+\rmi d_4$\\
  \hline $\frac{\Delta}{\sqrt{3}} (0,0,1,\xi_3,\xi_3^2)$ & $\Delta^2$ & 0 & $1$  & $4/3$ & \\
 \hline $\frac{\Delta}{2} (0,0,1,1,\sqrt{2}\rmi)$ & $\Delta^2$ & 0 & $1$  & $9/4$ & \\
  \hline $\frac{\Delta}{\sqrt{3}} (1,\xi_3^2,\xi_3,0,0)$ & $\Delta^2$ & 0 & $1.496$ & $3.654$ & \\
 \hline $\frac{\Delta}{\sqrt{5}} (1,\xi_5,\xi_5^2,\xi_5^3,\xi_5^4)$ & $\Delta^2$ & 0 & $1.527$ & $3.369$ & \\
 \hline $\frac{\Delta}{\sqrt{2}}(a,\rmi,\sqrt{1-a^2},0,0)$ & $\Delta^2$ & 0 & $\frac{2(9+\sqrt{6})}{15}$ & $\frac{8(9+\sqrt{6})}{25}$ & \\
\hline
\end{tabular}
\caption{Selection of complex tensor orders relevant for this work. The components of $\vec{\Delta}$ are related to the tensor $\phi$ via $\phi_{ij}=\Delta_aM^a_{ij}$, see Eq. (\ref{eq5}). The real orders satisfy $|\vec{\Delta}|^2=\vec{\Delta}^2$, whereas configurations with $\vec{\Delta}^2=0$ are genuinely complex. We define $\xi_n=e^{\rmi \pi/n}$ and $a=\sqrt{\frac{7-2\sqrt{6}}{15}}$. For each order parameter we display the quartic and sextic $\text{SO}(3)$ invariants $\mathcal{Q}=\frac{1}{2}I_8$ and $ \mathcal{S} = 9I_6I_7+I_4I_5$. In Luttinger semimetals, the d-wave superconducting equilibrium state just below the second-order phase transition is such that $\mathcal{S}$ needs to be maximized within the real or complex manifolds.}
\label{TabDelta}
\end{table}
\renewcommand{\arraystretch}{1}

\paragraph{From Luttinger semimetal to its GL theory.} After these general remarks we now turn to the particular realization of complex tensor order in three-dimensional electronic systems with the chemical potential $\mu$ close to an isotropic QBT point. The low-energy physics is assumed to be captured by the Lagrangian for interacting Luttinger fermions \cite{abrikosov, moon, PhysRevLett.113.106401}
\begin{align}
 \label{eq8} L= \psi^\dagger (\partial_\tau+d_a(\textbf{p})\gamma_a-\mu)\psi + g_1 (\psi^\dagger\psi)^2 + g_2 (\psi^\dagger\gamma_a\psi)^2,
\end{align}
which displays particle-hole, rotation, inversion, and time-reversal symmetry.
In $L$, $\psi$ is a four-component Grassmann field, $\tau$ denotes imaginary time, $\textbf{p}=-\rmi \nabla$ is the momentum operator, five $4\times 4$ matrices $\gamma_a$ satisfy Clifford algebra, $\{\gamma_a,\gamma_b\}=2\delta_{ab}$, summation over $a=1,\dots,5$ is implied, and the quadratic momentum dependence is captured by the $\ell=2$ spherical harmonics $d_a(\textbf{p})=\frac{\sqrt{3}}{2}p_ip_jM^a_{ij}$. In our units $\hbar=k_{\rm B}=2m_*=1$ with $m^*$ the effective electron mass. We choose the matrices $\gamma_{1,2,3}$ to be real and $\gamma_{4,5}$ to be complex, so that the time-reversal operator is given by $\mathcal{T}=\gamma_{45}\mathcal{K}$, where $\gamma_{ab}=\rmi \gamma_a\gamma_b$ and $\mathcal{K}$ denotes complex conjugation \cite{herbut2012,PhysRevB.93.205138}. It is assumed that the QBT point captures the band structure of an underlying material for momenta below the ultraviolet cutoff $\Lambda$.

The interaction terms in $L$ constitute a full (Fierz-complete) set of short-range interactions \cite{PhysRevLett.113.106401} in presence of rotational symmetry. Further local interactions inevitably contain powers of momenta and are thus suppressed for small $\mu$. We neglect the long-range part of the Coulomb interactions here \cite{abrikosov, moon, PhysRevLett.113.106401,JanssenHerbut}, which, although not screened, is assumed suppressed by either a large dielectric constant and/or a small effective electron mass. The interaction part of $L$ can be exactly rewritten as $L_{\rm s}+L_{\rm d}$ with  \cite{PhysRevB.93.205138}
\begin{align}
 \label{eq9} L_{\rm s} &=  g_{\rm s} (\psi^\dagger\gamma_{45}\psi^*)(\psi^{\rm T}\gamma_{45}\psi),\\
 \label{eq10} L_{\rm d} &= g_{\rm d} (\psi^\dagger \gamma_a\gamma_{45}\psi^*)(\psi^{\rm T}\gamma_{45}\gamma_a\psi),
\end{align}
where a nonvanishing expectation value $\Delta_{\rm s}=\langle \psi^{\rm T}\gamma_{45}\psi\rangle$ or $\Delta_a=\langle\psi^{\rm T} \gamma_{45}\gamma_a\psi\rangle$ would signal the onset of s- or d-wave superconductivity, respectively. The corresponding coupling constants are related to $g_{1,2}$ according to
\begin{align}
 \label{eq11} g_{\rm s} &= \frac{1}{4}(g_1+5g_2),\\
 \label{eq12} g_{\rm d} &= \frac{1}{4}(g_1-3g_2).
\end{align}
Crucially, an attraction in the d-wave pairing channel can be induced by a sufficiently large positive $g_2$, which, in addition, suppresses s-wave superconductivity.  This scenario is particularly appealing for YPtBi, where conventional electron-phonon-coupling cannot account for the large value of $T_{\rm c}$ \cite{PhysRevLett.116.137001}. We emphasize that $L_{\rm s,d}$ comprise local Cooper pairing and the angular dependence of the associated Cooper pair wave functions is trivial \cite{SOM}. In the following we assume $g_{\rm d}=-g<0$ and neglect $L_{\rm s}$. Despite its apparent five-component structure, the complex order parameter $\vec{\Delta}$ constitutes a representation of $\text{SO}(3)$; its entries transform under rotations as components of a second-rank tensor $\phi$ by means of Eq. (\ref{eq5}) \cite{SOM}.

The mean-field GL free energy $F(\phi)=F(\phi,T,\mu,\Lambda)$ for complex d-wave order to sextic order is given by
\begin{align}
 \nonumber F(\phi) &= r(g) |\vec{\Delta}|^2 + q_1 |\vec{\Delta}|^4 + q_2(\vec{\Delta}^2)(\vec{\Delta}^2)^*+s_1|\vec{\Delta}|^6\\
 \label{eq13} &+s_2(\vec{\Delta}^2)(\vec{\Delta}^2)^*|\vec{\Delta}|^2+s_3\mathcal{S}+\mathcal{O}(\phi^8)
\end{align}
with
\begin{align}
 \label{eq13b} \mathcal{S} = 9|\mbox{tr}(\phi^2\phi^\dagger)|^2+|\mbox{tr}(\phi^3)|^2.
\end{align}
The expressions for the coefficients are given in the supplemental material (SM) \cite{SOM}. Remarkably, not all symmetry-allowed invariant combinations from Eqs. (\ref{eq2})-(\ref{eq4}) appear in the one-loop result. Especially, the quartic invariant $\mathcal{Q}=\frac{1}{2}I_8$ does not show up to sextic order. 
Consequently, since the free energy to quartic order only depends on the invariants $I_1$ and $I_2I_3$ as in Eq. (\ref{eq7}), the quartic theory has an accidental $\text{SO}(5)\times\text{U}(1)$ symmetry \cite{PhysRevA.9.868}.  Therefore, the energetically most favorable configuration is selected by the higher-order terms beyond the quartic level, which reduce the symmetry to the physical $\text{SO}(3)\times\text{U}(1)$; close to a second-order phase transition these are the sextic terms. Crucially, $s_3<0$  \cite{SOM}, so that $\mathcal{S}$ needs to be {\it maximized} within the real or complex manifolds.

\begin{figure}[t]
\centering
\includegraphics[width=8.5cm]{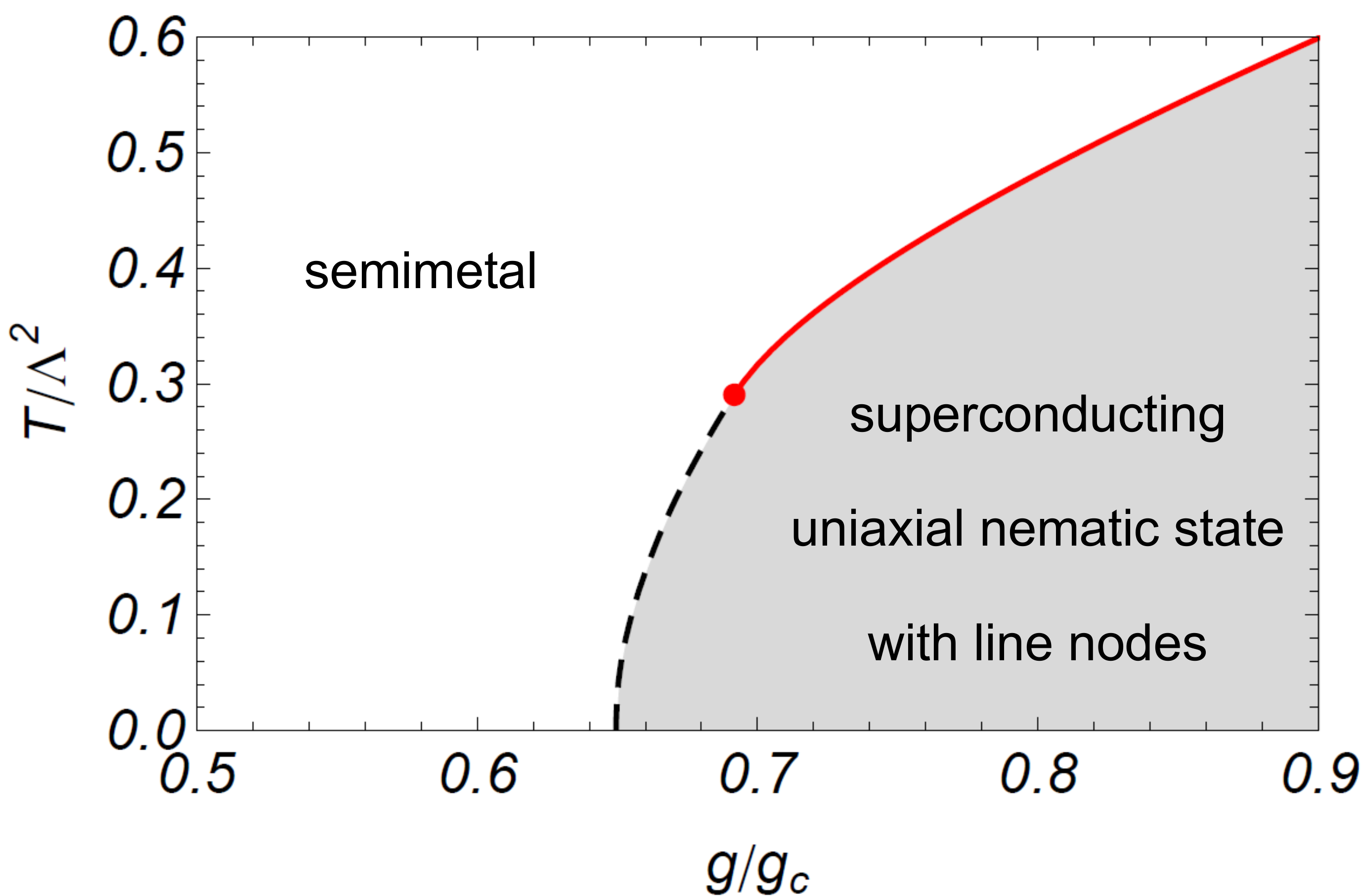}
\caption{Mean-field phase structure of unconventional d-wave superconductivity in the limit $\mu\to0$. The phase transition into the uniaxial nematic state featuring line nodes in the spectrum requires a sufficiently large coupling $g$, where $g_{\rm c}=\frac{10\pi^2}{\Lambda}$ is a reference coupling. The solid (red) and dashed (black) lines indicate second- and first-order transitions, respectively, meeting at a tricritical point (red dot).}
\label{FigZeroMu}
\end{figure}

\paragraph{Phase diagrams and superconducting states.} We first discuss the limiting case of the mean-field phase diagram for $\mu=0$, shown in Fig. \ref{FigZeroMu}. Besides temperature, $T$, the only energy scale present in the problem is the ultraviolet cutoff $\Lambda^2$. The coupling constant $g$ is naturally parametrized in terms of the critical coupling for a putative quantum critical point for d-wave order \cite{PhysRevB.93.205138}, given by $g_{\rm c}= \frac{10\pi^2}{\Lambda}$ within our regularization scheme.
For sufficiently high $T/\Lambda^2$, the transition is of second-order and thus described well by the expansion in Eq. (\ref{eq13}). In this regime $q_2<0$ and $q_1 + q_2 >0$,  so that real order develops upon increasing $g$. The sextic term selects the uniaxial nematic state $\vec{\Delta}=\Delta(0,1,0,0,0)$ as the state of maximal $\mathcal{S}$. The line of second-order transition terminates at a tricritical point $(g/g_{\rm c},T/\Lambda^2)=(0.69,0.29)$, where the combination $q_1+q_2$ changes sign and the transition consequently becomes first-order. To estimate the first-order line, we compute the non-expanded function $F(\phi)$ at the mean-field level. We find that the uniaxial nematic state has the lowest free energy among the real and complex solutions. The transition for $T/\Lambda^2\to 0$ occurs at $g/g_{\rm c}=0.65$.

The uniaxial nematic state that emerges here features line nodes in the excitation spectrum of quasiparticles. Indeed, the spectrum for real $\vec{\Delta}$ is given by
\begin{align}
 \label{eq15} |E_{\pm}(\textbf{p})| &= \sqrt{|\textbf{p}|^4+p_0^4\pm 2\sqrt{|\textbf{p}|^4p_0^4-[\vec{d}(\textbf{p})\cdot\vec{\Delta}]^2}},
\end{align}
with $p_0=(\Delta^2+\mu^2)^{1/4}$, so that line nodes are determined by the single condition $\vec{d}(\textbf{p})\cdot\vec{\Delta}=0$; see the discussion and plots in the SM \cite{SOM} for general real $\vec{\Delta}$. For the uniaxial nematic state the nodes are along two parallel circles on a momentum sphere of radius $p_0$.

In half-Heuslers, the chemical potential $\mu\neq 0$ and, despite exceptionally low carrier densities, typically $T\ll \mu$. The nonzero value of $\mu$ implies a BCS-like instability for arbitrarily weak coupling $g>0$. This leads to a second-order phase transition with the critical temperature
\begin{align}
 \label{eq16} \frac{T_{\rm c}(g)}{\mu} =  \frac{8e^{\gamma+\frac{2}{3}-\frac{\pi}{2}}}{\pi} \exp\Bigl\{-\frac{1-g/g_{\rm c}}{g/g_{\rm c}} \frac{2}{\sqrt{\mu/\Lambda^2}}\Bigr\}
\end{align}
for small $g$ \cite{SOM}. Here $\gamma$ is Euler's constant and the numerical prefactor of the exponential is 1.8. Since experimental results are likely to be in the regime of sufficiently weak $g$, we can use this formula to estimate $g/g_{\rm c}$. For instance, from the experimental data reported in Ref. \cite{2016arXiv160303375K} on YPtBi we estimate $T_{\rm c}/\mu\simeq0.002$ and $\mu/\Lambda^2\simeq0.4$ so that $g/g_{\rm c}\simeq 0.3$. In fact, this coupling is surprisingly large and puts the half-Heusler materials within the range of physics discussed in this work.

\begin{figure}[t]
\centering
\includegraphics[width=8.5cm]{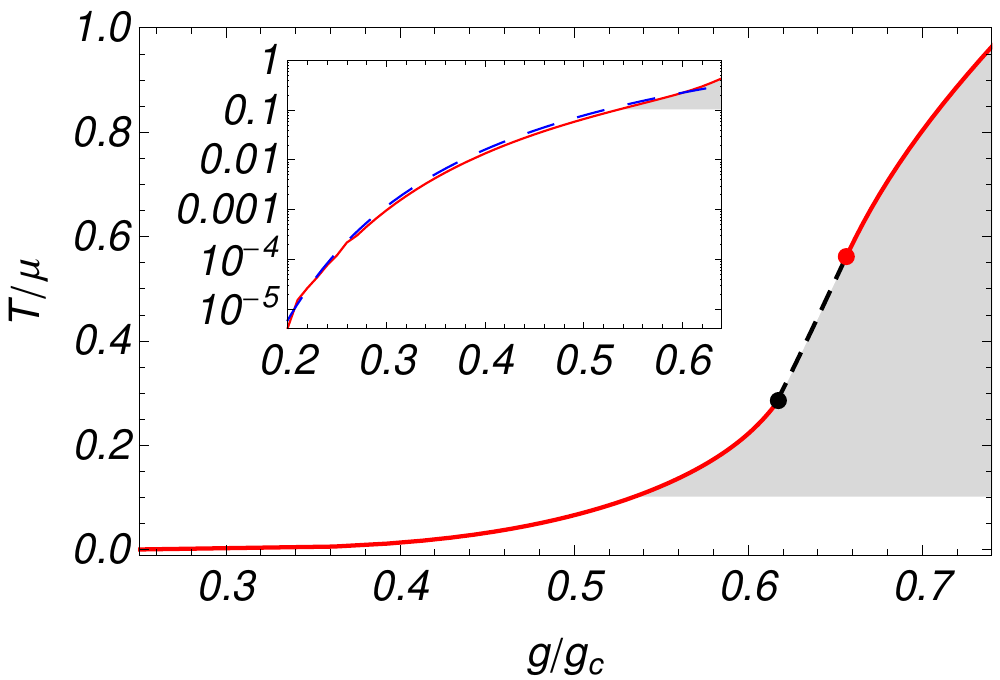}
\caption{Phase structure for $\mu>0$. As in Fig. \ref{FigZeroMu}, the solid (red) and dashed (black) lines indicate second- and first-order phase transitions. The limit $\mu\to 0$ can qualitatively be understood by pushing the lower second-order line towards the abscissa. This second-order transition line at lower $T/\mu$ is captured well by the weak coupling formula in Eq. (\ref{eq16}), as highlighted in the inset, where the long-dashed (blue) line is the weak-coupling result. Along the second-order lines, the transition is into the uniaxial nematic state for $T/\mu\geq 0.10$ (gray shaded region), and into a complex order with $\vec{\Delta}^2=0$ for $T/\mu\leq 0.10$. In the plot $\mu/\Lambda^2=0.4$.}
\label{FigFiniteMu}
\end{figure}

The phase structure for nonzero chemical potential depends on whether $\mu$ is in the range $\mu\ll \Lambda^2$ or $\mu \lesssim \Lambda^2$. For sufficiently small $\hat{\mu}=\mu/\Lambda^2$ the qualitative phase structure is as in Fig. \ref{FigFiniteMu}, where we have chosen $\hat{\mu}=0.4$ for illustration. The small-$\mu$ transition into the uniaxial nematic state appears at higher values of $T/\mu$, whereas the BCS-like transition described by Eq. (\ref{eq16}) appears for exponentially small $T/\mu$. Along the BCS-like line for $T/\mu\leq 0.10$ the transition occurs into a complex ordered state with maximal $\mathcal{S}$, whereas for $T/\mu>0.10$ uniaxial nematic order is selected. For larger $T/\mu$ this consistently connects with the phase structure obtained in the limit of $\mu=0$. The length of the first-order line connecting the two second-order lines in Fig. \ref{FigFiniteMu} shrinks with increasing $\hat{\mu}$. In particular, within our approximation it disappears for $\hat{\mu}>0.45$, and all transitions become then of second order. From our analysis we cannot exclude that further first-order transitions occur within the superconducting region (that is for larger $g$).

\paragraph{Symmetry-reducing perturbations.} The most general QBT point is described by the Luttinger Hamiltonian which also features a particle-hole asymmetry and cubic anisotropy, quantified through parameters $x$ and $\delta$, respectively \cite{PhysRevB.95.075149}. For the half-Heusler material YPtBi we estimate $x=0.17$ and $\delta=-0.19$ \cite{SOM}, which is small compared to the prefactor of unity of the term $d_a\gamma_a$ in $L$. The high-symmetry ansatz considered here thus provides a computationally efficient way to approach the phase structure and free energy within a few tens of percent. Furthermore, terms that break inversion symmetry in YPtBi are of the order of $<1\%$ for realistic values of $\mu/\Lambda^2$. The observation of line nodes thus disagrees with the finding of a complex ground state at low temperatures \cite{PhysRevLett.118.127001,SOM}. Higher-order corrections in $g$ or additional first-order transitions could, however, favor the uniaxial nematic state featuring line nodes even for $T/\mu\leq 0.10$.

An intriguing question arises from the presence of the anisotropy parameter $\delta\neq 0$ that restricts rotations to the cubic symmetry group.
We say that $\phi$ belongs to the representation E or T$_2$ if there is a frame such that $\phi=\phi_E=\sum_{a=1,2}\Delta_aM^a$ or $\phi=\phi_T=\sum_{a=3,4,5}\Delta_aM^a$, respectively, which remains invariant under cubic transformations. The largest values of $\mathcal{S}$ within E and T$_2$ are achieved for physically inequivalent configurations $\vec{\Delta}_E=\frac{\Delta}{\sqrt{2}}(1,\rmi,0,0,0)$ and $\vec{\Delta}_T=\frac{\Delta}{2}(0,0,1,1,\sqrt{2}\rmi)$, respectively. For the isotropic system, however,  there exist states with $\mathcal{S}/\Delta^6\geq \frac{8(9+\sqrt{6})}{25}=3.664$ (last entry in Tab. \ref{TabDelta}) which are neither entirely in E nor T$_2$. For nonzero $\delta$ the quadratic part of the free energy has the form $F_2(\delta) = r_{E}(\delta)\ \mbox{tr}(\phi_E^\dagger\phi_E)+r_{T}(\delta)\ \mbox{tr}(\phi_T^\dagger\phi_T)$ so that $\delta<0$ ($\delta>0$) penalizes the last three (first two) components of $\vec{\Delta}$. Within a certain region of small $|\delta|>0$ this penalty is small and the energetic gain from the admixture of E and T$_2$ in $\vec{\Delta}$ remains favorable.

\paragraph{Experiment and further directions.} Experimentally, the time-reversal symmetry breaking phase, being typically also magnetic, can be identified through muon spin resonance measurements. 
The line nodes of the nematic orders and their orientation result in characteristic temperature and directional dependences of electromagnetic and thermodynamic responses, or are accessible via ARPES. Experimental responses to strain of the nematic state in the E representation are discussed by Fu in Ref. \cite{PhysRevB.90.100509}.
From a theoretical perspective, a complete classification of complex orders with $\vec{\Delta}^2=0$ is desirable, and evaluation of the non-expanded free energy $F(\phi)$ beyond mean-field theory \cite{PhysRevA.91.013610,BoettcherSarma} becomes mandatory to find the equilibrium state. The interplay with s-wave superconductivity and magnetism, and the influence of antisymmetric spin-orbit coupling, constitute promising routes towards novel phases of matter.

\acknowledgements We gratefully acknowledge useful discussions with L. Classen and M. M. Scherer.  IB acknowledges funding by the DFG under Grant No. BO 4640/1-1. IFH is supported by the NSERC of Canada.

\bibliographystyle{apsrev4-1}
\bibliography{refs_scnem}

\cleardoublepage

\setcounter{equation}{0}
\renewcommand{\theequation}{S\arabic{equation}}

\begin{center}
\textbf{\Large Supplemental Material}
\end{center}

\section{Tensor order and Cooper pairing}\label{AppCoop}
We give concrete expressions for the complex order parameters $\phi$ in terms of the spin-3/2 electronic degrees of freedom. Electrons are parametrized by the four-component spinor
\begin{align}
 \label{coop0} \psi = \begin{pmatrix} \psi_1 \\ \psi_2 \\ \psi_3 \\ \psi_4 \end{pmatrix} = \begin{pmatrix} c_{3/2} \\ c_{1/2} \\ c_{-1/2} \\ c_{-3/2} \end{pmatrix}.
\end{align}
In a Heisenberg picture, $c_{m_j}$ is replaced by the annihilation operator for an electron in angular momentum state $|j=3/2,m_j\rangle$. We choose the standard representation of spin-3/2 angular momentum matrices $J_i$ where $J_z=\text{diag}(3/2,1/2,-1/2,-3/2)$ is diagonal, which also implies the identification Eq. (\ref{coop0}).
We construct the symmetric traceless second rank tensor $S_{ij}=J_iJ_j+J_jJ_i-\frac{5}{2}\delta_{ij}\mathbb{1}_4$, with $4\times 4$ unit matrix $\mathbb{1}_4$. The anti-commuting matrices $\gamma_a$ are then defined by
\begin{align}
 \label{coop4} \gamma_a = \frac{1}{2\sqrt{3}} S_{ij}M^a_{ij},
\end{align}
see Ref. \cite{PhysRevB.95.075149} for a detailed account. With the particular representation chosen for $J_i$ we obtain
\begin{align}
 \label{coop5} \gamma_1 &= \begin{pmatrix}0 & 0 & 1 & 0 \\ 0 & 0 & 0 & 1 \\ 1 & 0 & 0 & 0 \\ 0 & 1 & 0 & 0 \end{pmatrix},\ \gamma_2 = \begin{pmatrix} 1 & 0 & 0 & 0 \\ 0 & -1 & 0 & 0 \\ 0 & 0 & -1 & 0 \\ 0 & 0 & 0 & 1 \end{pmatrix},\\
 \label{coop6} \gamma_3 &= \begin{pmatrix} 0 & 1 & 0 & 0 \\ 1 & 0 & 0 & 0 \\ 0 & 0 & 0 & -1 \\ 0 & 0 & -1 & 0 \end{pmatrix},\ \gamma_4 = \begin{pmatrix}0 & -\rmi & 0 & 0 \\ \rmi & 0 & 0 & 0 \\0 & 0 & 0 & \rmi \\  0 & 0 & -\rmi & 0 \end{pmatrix},\\
 \label{coop7} \gamma_5 &= \begin{pmatrix}0 & 0 & -\rmi & 0 \\ 0 & 0 & 0 & -\rmi \\ \rmi & 0 & 0 & 0 \\ 0 & \rmi & 0 & 0 \end{pmatrix}.
\end{align}
The components $\Delta_a$ of the complex order parameter $\phi$ are given by $\Delta_a=\langle\psi^{\rm T}\gamma_{45}\gamma_a\psi\rangle$. This eventually yields
\begin{align}
 \label{coop8} \Delta_1 &= -2\rmi \langle c_{3/2}c_{1/2}+c_{-1/2}c_{-3/2}\rangle,\\
 \label{coop9} \Delta_2 &=-2\rmi \langle c_{3/2}c_{-3/2}+c_{1/2}c_{-1/2}\rangle,\\
 \label{coop10} \Delta_3 &= 2\rmi \langle c_{3/2}c_{-1/2}-c_{1/2}c_{-3/2}\rangle ,\\
 \label{coop11} \Delta_4 &= -2\langle c_{3/2}c_{-1/2}+c_{1/2}c_{-3/2}\rangle ,\\
 \label{coop12} \Delta_5 &=2 \langle c_{3/2}c_{1/2}-c_{-1/2}c_{-3/2}\rangle.
\end{align}
These expressions are consistent with the expressions for quintet pairing given in Ref. \cite{PhysRevLett.116.177001}, with $(\Delta_1,\Delta_2)$ and $(\Delta_3,\Delta_4,\Delta_5)$ belonging to the E- and T$_2$-representations, respectively. For completeness we also display the s-wave order parameter, which reads
\begin{align}
 \label{coop13} \Delta_{\rm s} = \langle\psi^{\rm T}\gamma_{45}\psi\rangle = -2\rmi \langle c_{3/2}c_{-3/2}-c_{1/2}c_{-1/2}\rangle.
\end{align}
Using the inverse relation to Eq. (\ref{coop4}), namely $S_{ij}=\sqrt{3}\gamma_aM^a_{ij}$, we obtain
\begin{align}
 \label{coop14} \phi_{ij} = \Delta_aM^a_{ij} = \frac{1}{\sqrt{3}} \langle \psi^{\rm T}\gamma_{45} S_{ij}\psi\rangle,
\end{align}
which underlines again that $\phi$ transforms as an irreducible second-rank tensor under rotations.

\section{Luttinger Hamiltonian and phenomenological parameters for YPtBi}\label{AppLut}
We relate our isotropic model to the more general Luttinger Hamiltonian and discuss phenomenological parameters for the band structure of YPtBi to estimate how much they deviate from the more symmetric ansatz considered in the main text. The isotropic Luttinger Hamiltonian is given by
\begin{align}
  \nonumber H &= \Bigl(\alpha_1+\frac{5}{2}\alpha_2\Bigr)p^2\mathbb{1}_4-2\alpha_3(\textbf{p}\cdot\vec{J})^2+2(\alpha_3-\alpha_2)\sum_{i=1}^3p_i^2J_i^2\\
 \nonumber &= \Bigl(\alpha_1+\frac{5}{2}\alpha_2\Bigr)p^2\mathbb{1}_4-2\alpha_3\sum_{i\neq j}p_ip_jJ_iJ_j-2\alpha_2\sum_{i=1}^3p_i^2J_i^2\\
  \label{lut1} &= \alpha_1 p^2\mathbb{1}_4 -(\alpha_2+\alpha_3)\sum_{a=1}^5 d_a\gamma_a+(\alpha_2-\alpha_3)\sum_{a=1}^5 s_ad_a\gamma_a
\end{align}
with Luttinger parameters $\alpha_{1,2,3}$ and
\begin{align}
 d_1 &= \frac{\sqrt{3}}{2}(p_x^2-p_y^2),\ d_2 = \frac{1}{2}(2p_z^2-p_x^2-p_y^2),\\
 d_3 &= \sqrt{3}p_zp_x,\ d_4 = \sqrt{3}p_yp_z,\ d_5 = \sqrt{3}p_xp_y.
\end{align}
We define $s_{1,2}=-1$, $s_{3,4,5}=+1$. We normalize the field $\psi$ such that the coefficient in front of $d_a\gamma_a$ is unity in Eq. (\ref{lut1}) and write
\begin{align}
 \label{lut2} H = x p^2\mathbb{1}_4 +\sum_{a=1}^5(1+\delta s_a) d_a\gamma_a,
\end{align}
such that
\begin{align}
 \label{lut3} x = -\frac{\alpha_1}{\alpha_2+\alpha_3},\ \delta = -\frac{\alpha_2-\alpha_3}{\alpha_2+\alpha_3}
\end{align}
quantify the particle-hole asymmetry and cubic anisotropy, respectively. The isotropic model considered in the main text corresponds to $x=\delta=0$. The isotropic model is a good leading order approximation if $x,\delta$ are small compared to unity.

The phenomenological Hamiltonian for YPtBi given in Ref. \cite{2016arXiv160303375K} reads
\begin{align}
 \label{lut4} H = \alpha p^2\mathbb{1}_4 +\gamma \sum_{i\neq j} p_ip_jJ_iJ_j +\beta\sum_i p_i^2J_i^2+H_{\rm nc}
\end{align}
with $\alpha=20.5 \text{eV}a^2/\pi^2$, $\beta=-18.5\text{eV}a^2/\pi^2$, $\gamma=-12.7\text{eV}a^2/\pi^2$, $a$ the lattice constant, and $H_{\rm nc}$ a non-centrosymmetric contribution to be discussed below. Comparing with the second line of Eq. (\ref{lut1}) we deduce $\alpha=\frac{\hbar^2}{2m^*}(\alpha_1+5\alpha_2/2)$, $\gamma=\frac{\hbar^2}{2m^*}(-2\alpha_3)$, $\beta=\frac{\hbar^2}{2m^*}(-2\alpha_2)$. Consequently,
\begin{align}
 \label{lut5} x &= \frac{2(\alpha+5\beta/4)}{\gamma+\beta} = 0.17,\\
 \label{lut6} \delta &= \frac{\gamma-\beta}{\gamma+\beta} = -0.19.
\end{align}
The inversion symmetry breaking term $H_{\rm nc}$ is given by
\begin{align}
 \label{lut7} H_{\rm nc} = \hat{\delta} \sum_i p_i(J_{i+1}J_iJ_{i+1}-J_{i+2}J_iJ_{i+2})
\end{align}
with $\hat{\delta}=0.06\text{eV}a/\pi$. Since $H_{\rm nc}$ is linear in momenta, it will be the dominant contribution for small momenta. The typical momentum scale of excitations is given by the Fermi momentum $p_{\rm F}=\sqrt{\mu}$. The ultraviolet cutoff reads $\Lambda=\pi/a$. Consequently, the relative strength of $\hat{\delta}$ and $\alpha$ (representing a typical term of the quadratic Luttinger Hamiltonian) is determined by the ratio
\begin{align}
\label{lut8} R_{\rm nc} = \frac{\hat{\delta}p_{\rm F}}{\alpha p_{\rm F}^2} = \frac{\hat{\delta}}{\alpha\sqrt{\mu}} = \frac{0.06}{20.5} \frac{1}{\sqrt{\mu/\Lambda^2}}.
\end{align}
For generic values of $\mu/\Lambda^2$ we have $R_{\rm nc}\ll 1$. In order to have $R_{\rm nc}\sim 1$, the chemical potential needs to be as small as $\mu/\Lambda^2\sim10^{-5}$. From the fitting value $\mu=35\text{meV}$ from Ref. \cite{2016arXiv160303375K} together with the critical temperature $T_{\rm c}= 0.8 \text{K}$ for YPtBi we conclude $T_{\rm c}/\mu=0.002$. From Fig 2b of Ref. \cite{2016arXiv160303375K} we further estimate the ratio of $\mu$ over the bandwidth to be on the order of several tens percent, which motivates us to choose $\mu/\Lambda^2=0.4$ for our Fig. 2 in the main text.

\section{Excitation spectrum and line nodes}\label{AppNode}
Here we discuss nodes in the excitation spectrum for real and complex orders. In particular, we show that real tensor orders \emph{always} feature line nodes.

\begin{figure}[t!]
\centering
\begin{minipage}{8.5cm}
\includegraphics[width=2.7cm]{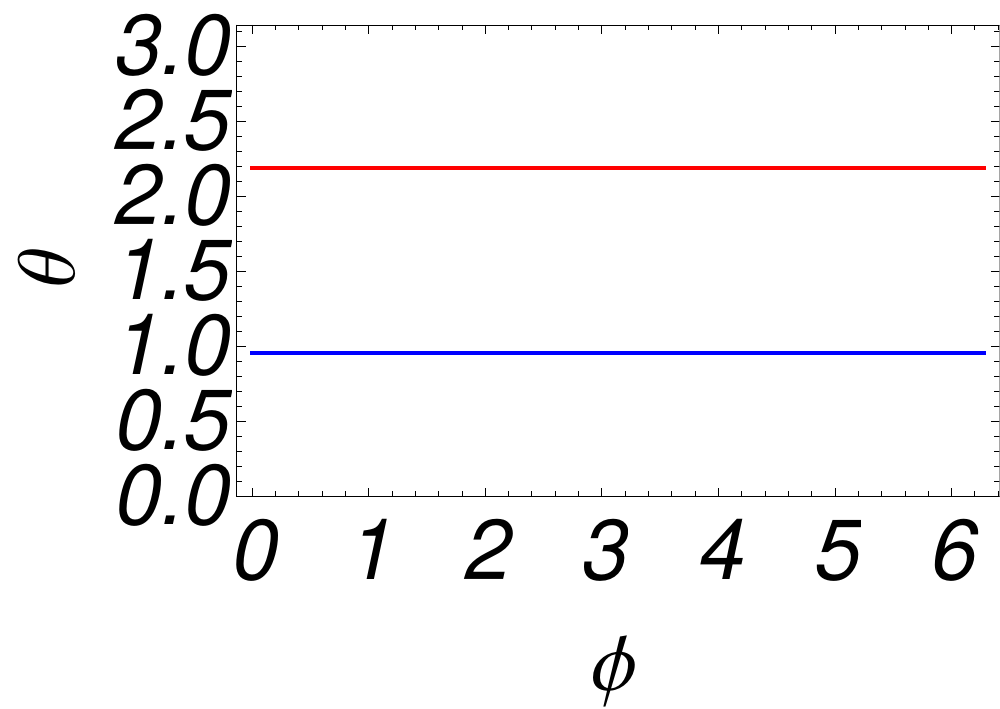}
\includegraphics[width=2.7cm]{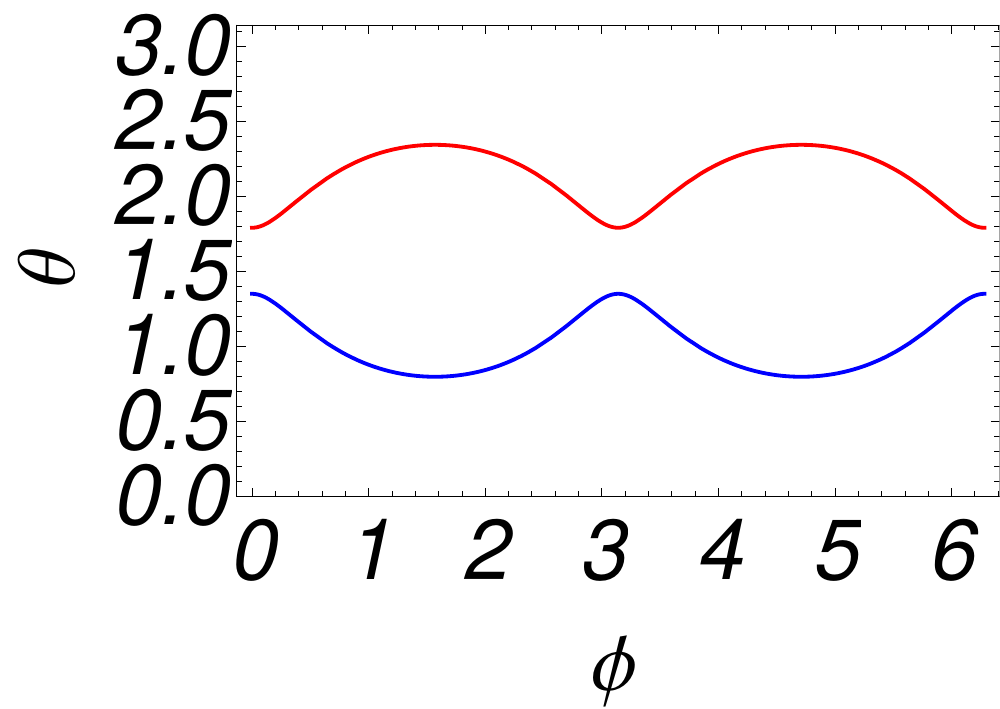}
\includegraphics[width=2.7cm]{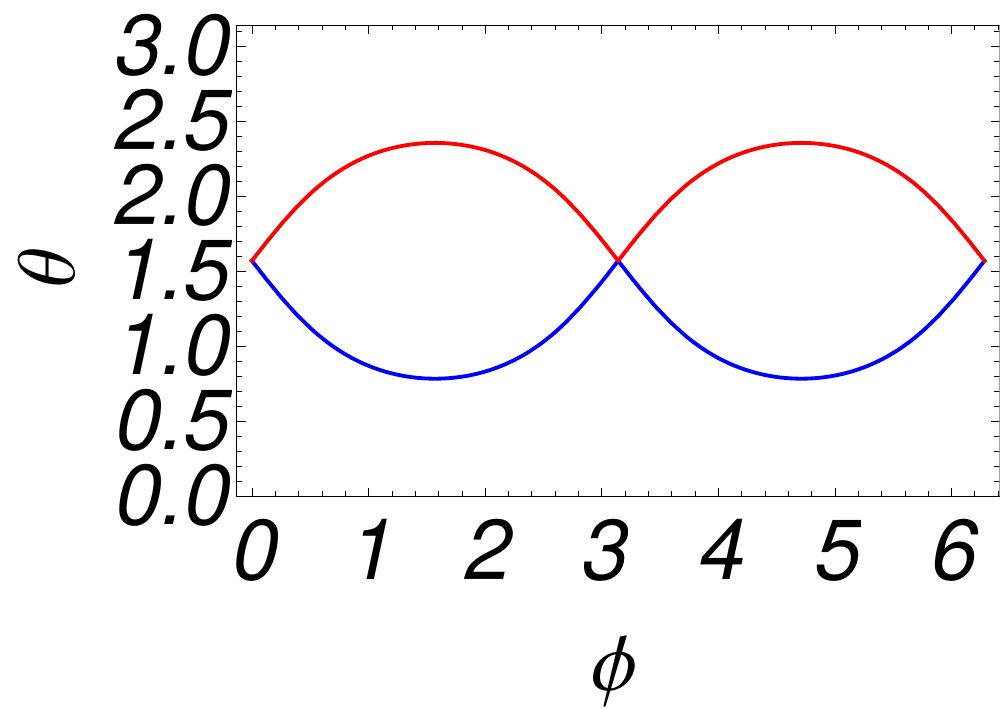}\\
\includegraphics[width=2.7cm]{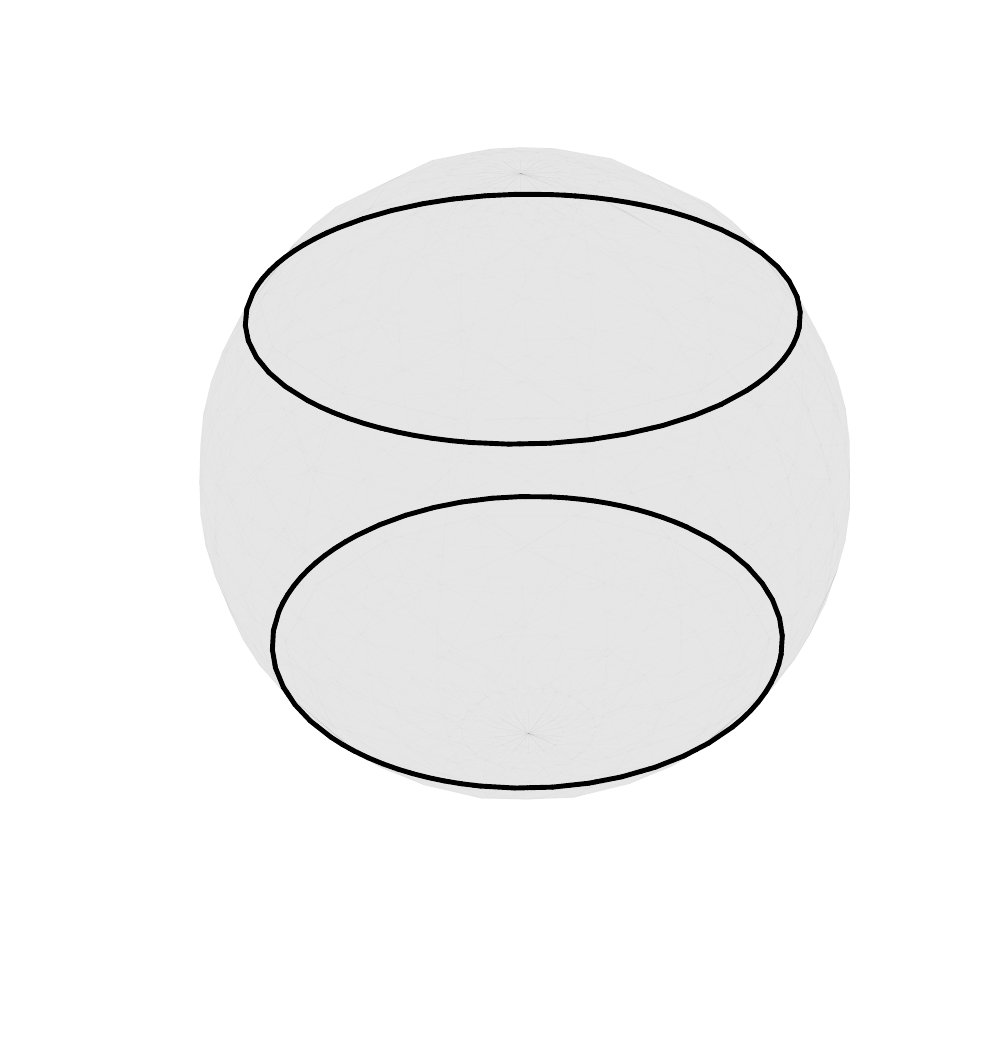}
\includegraphics[width=2.7cm]{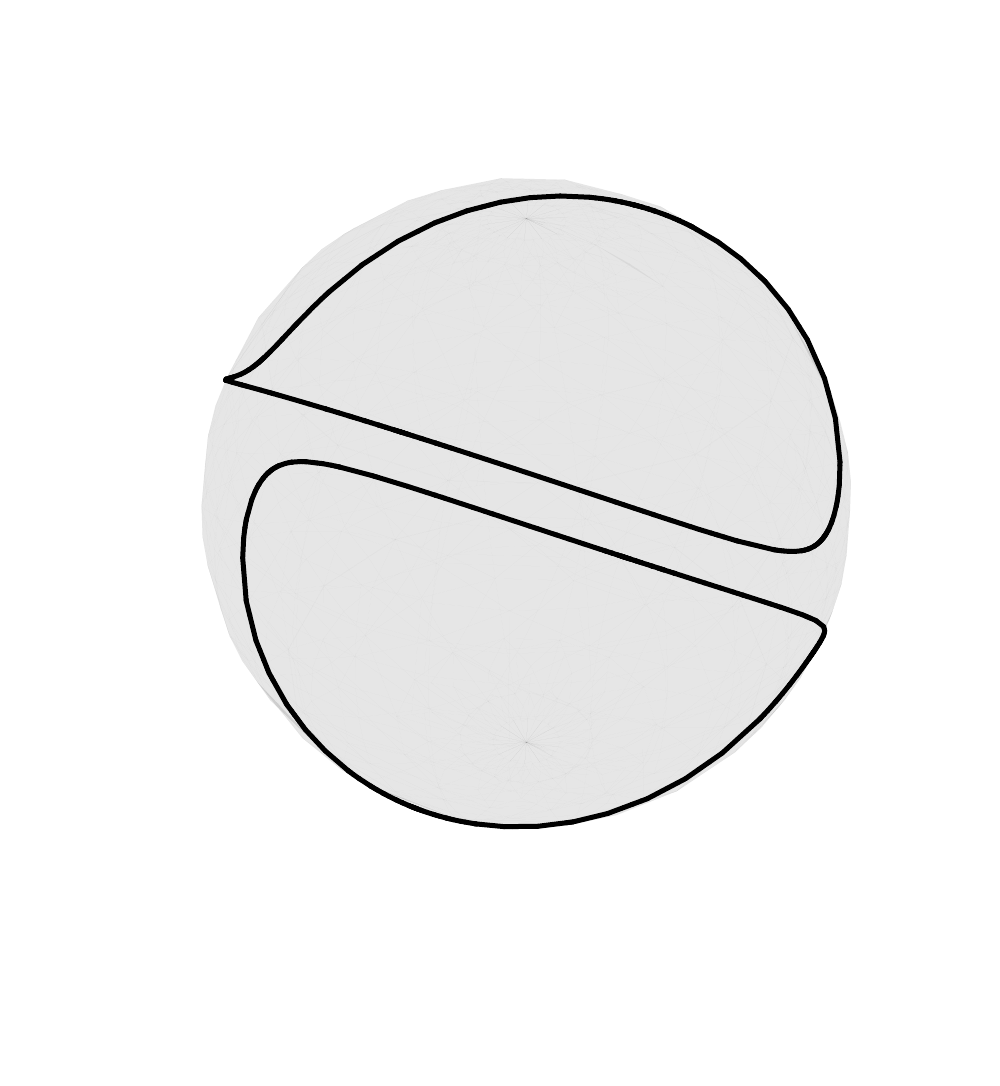}
\includegraphics[width=2.7cm]{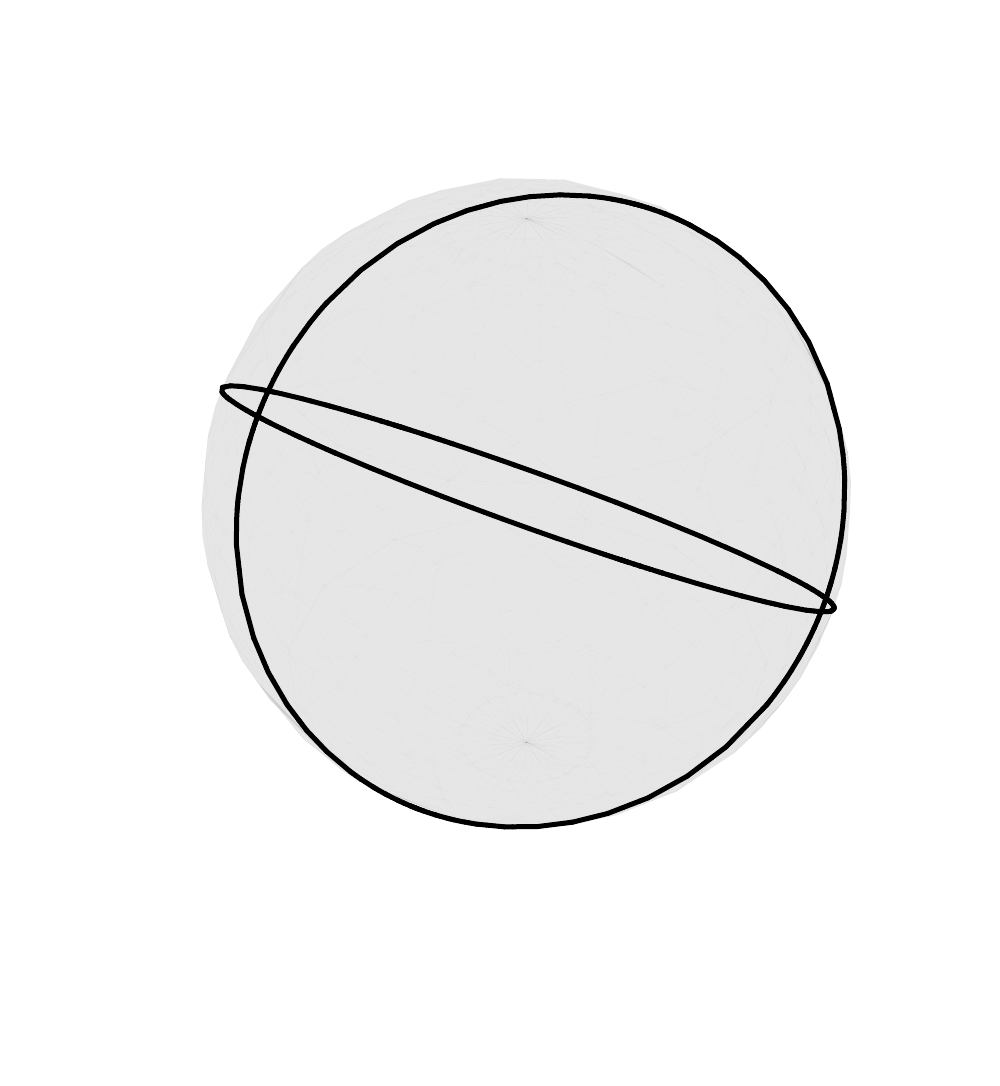}\\
\includegraphics[width=2.7cm]{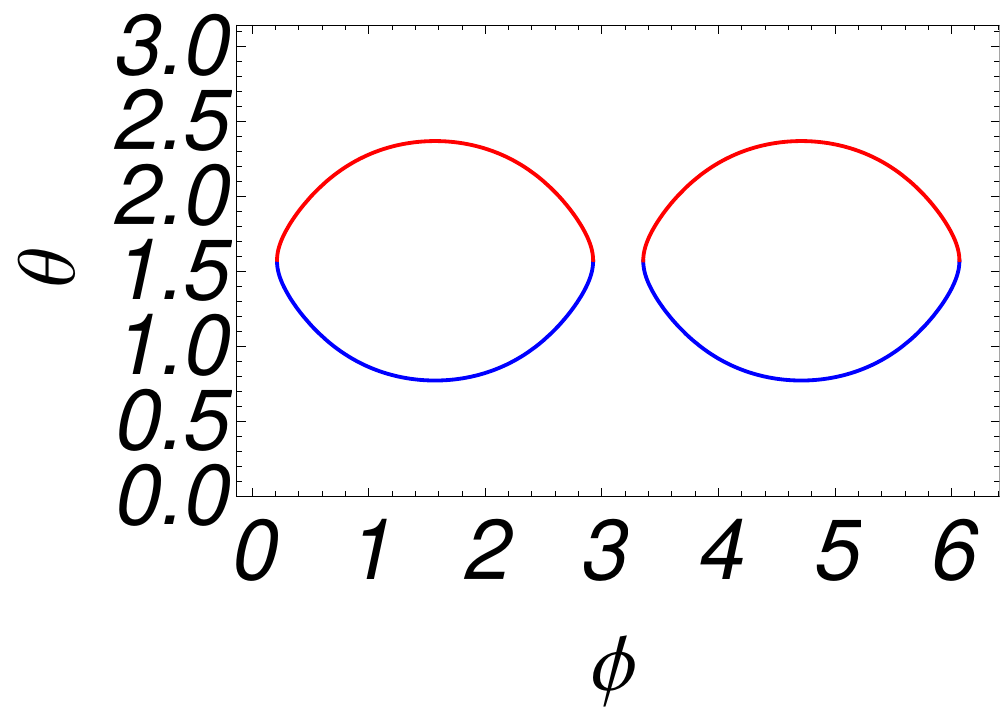}
\includegraphics[width=2.7cm]{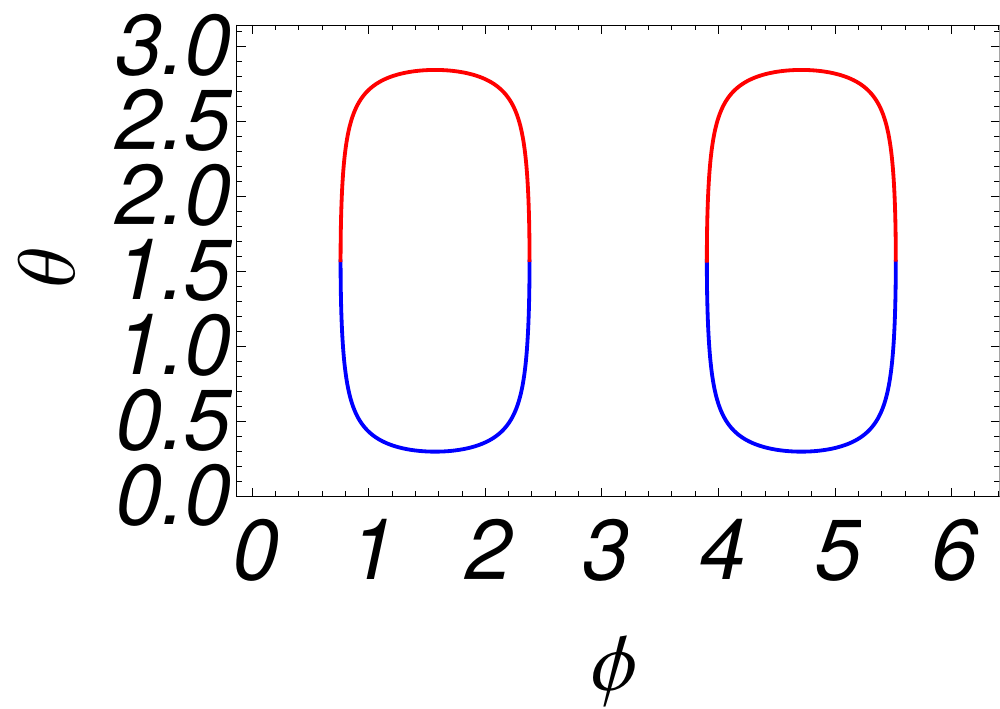}
\includegraphics[width=2.7cm]{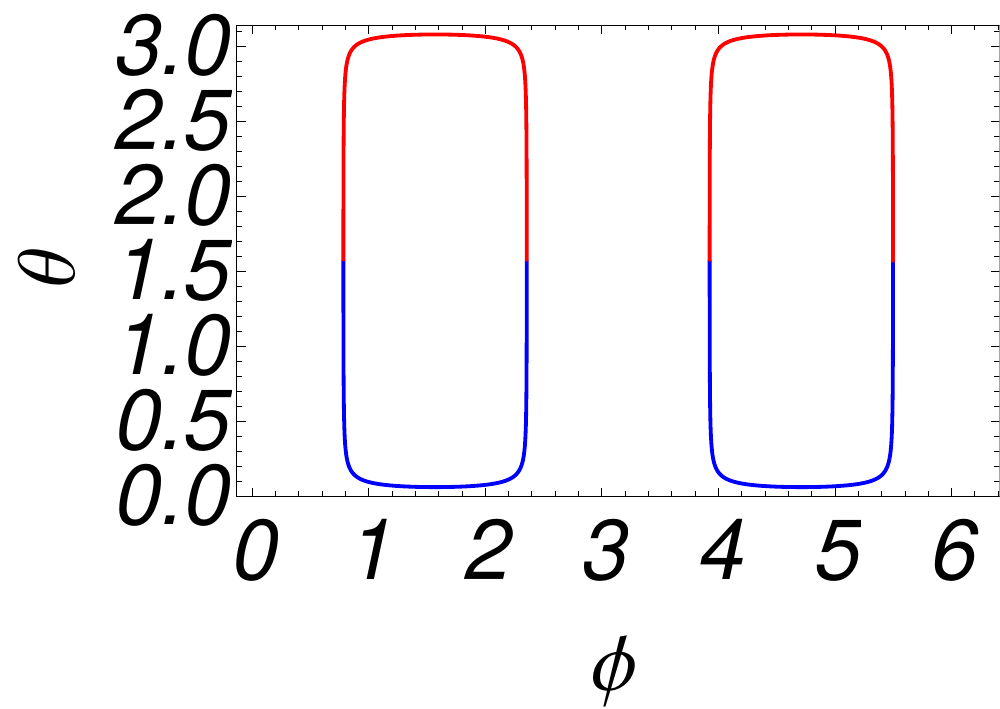}\\
\includegraphics[width=2.7cm]{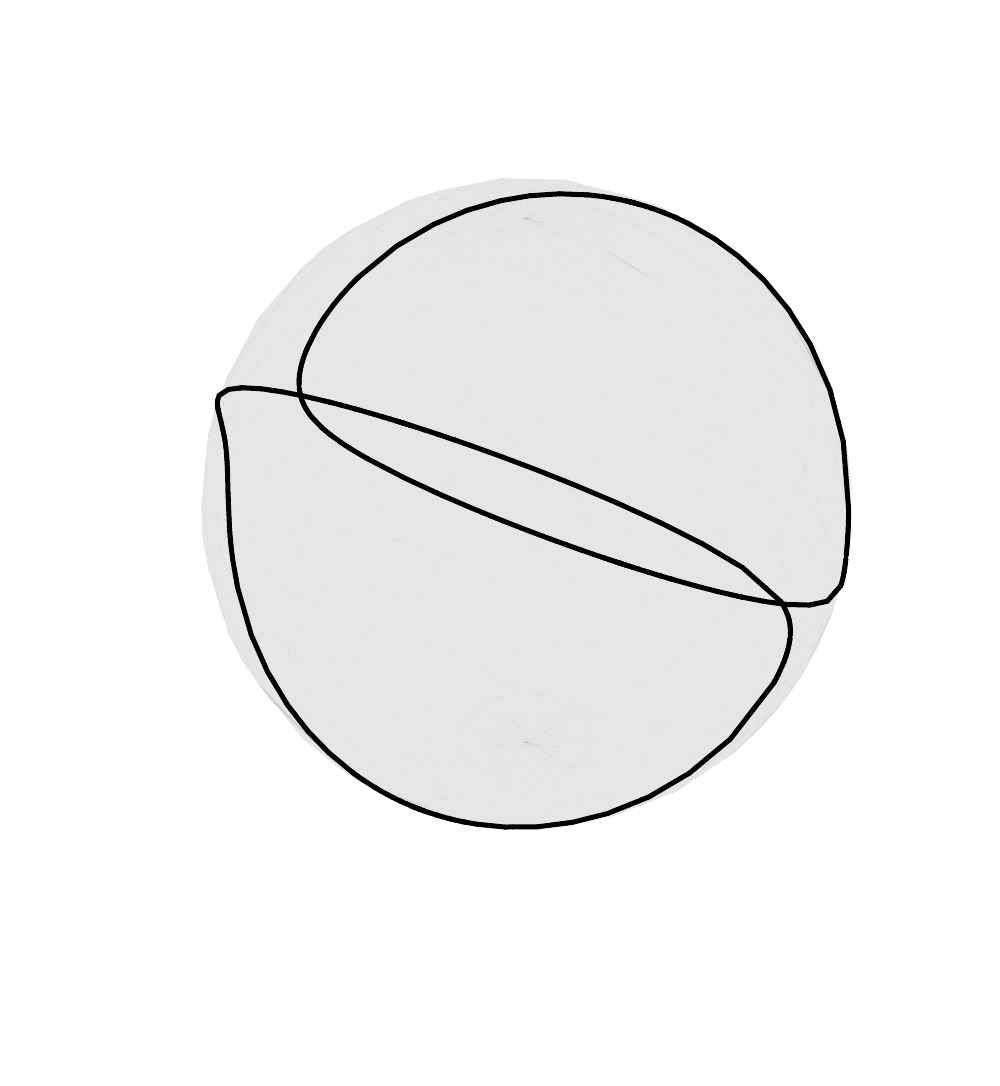}
\includegraphics[width=2.7cm]{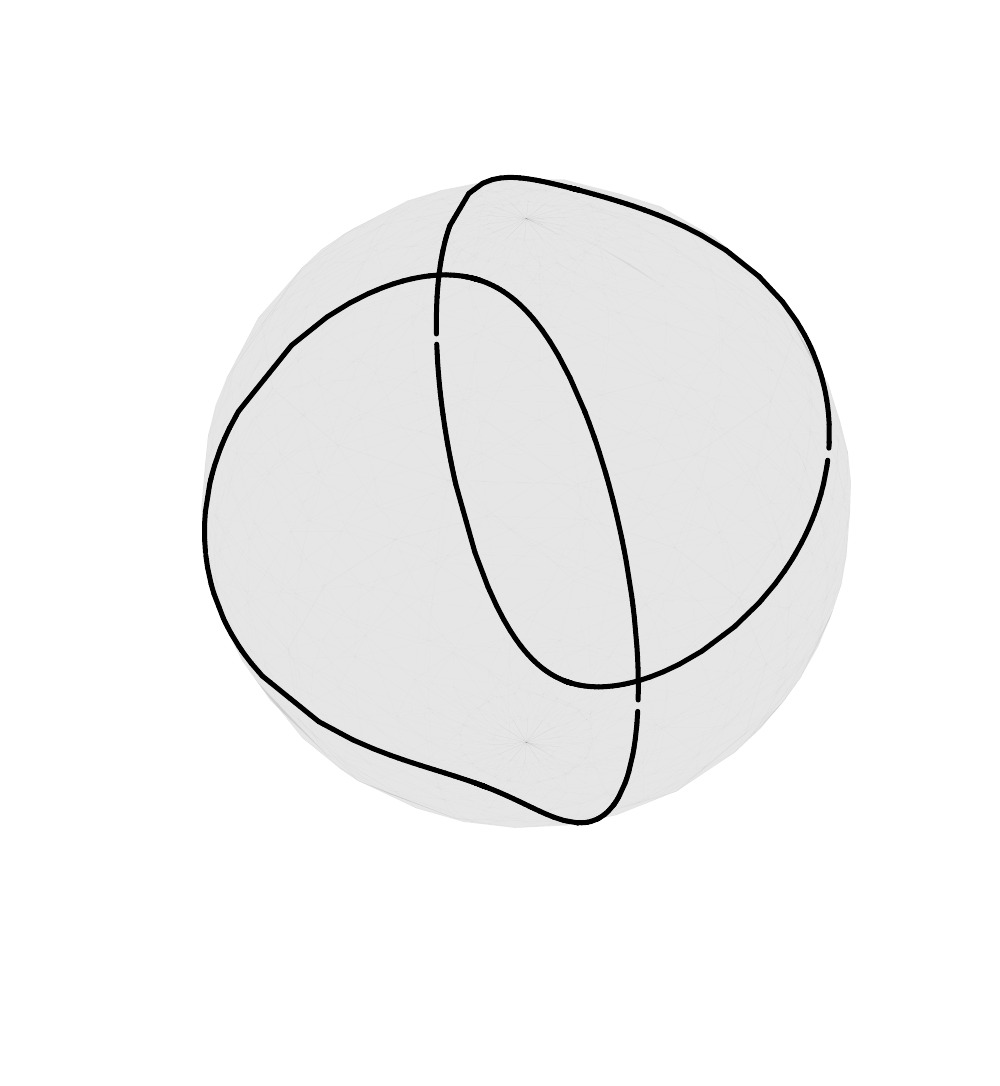}
\includegraphics[width=2.7cm]{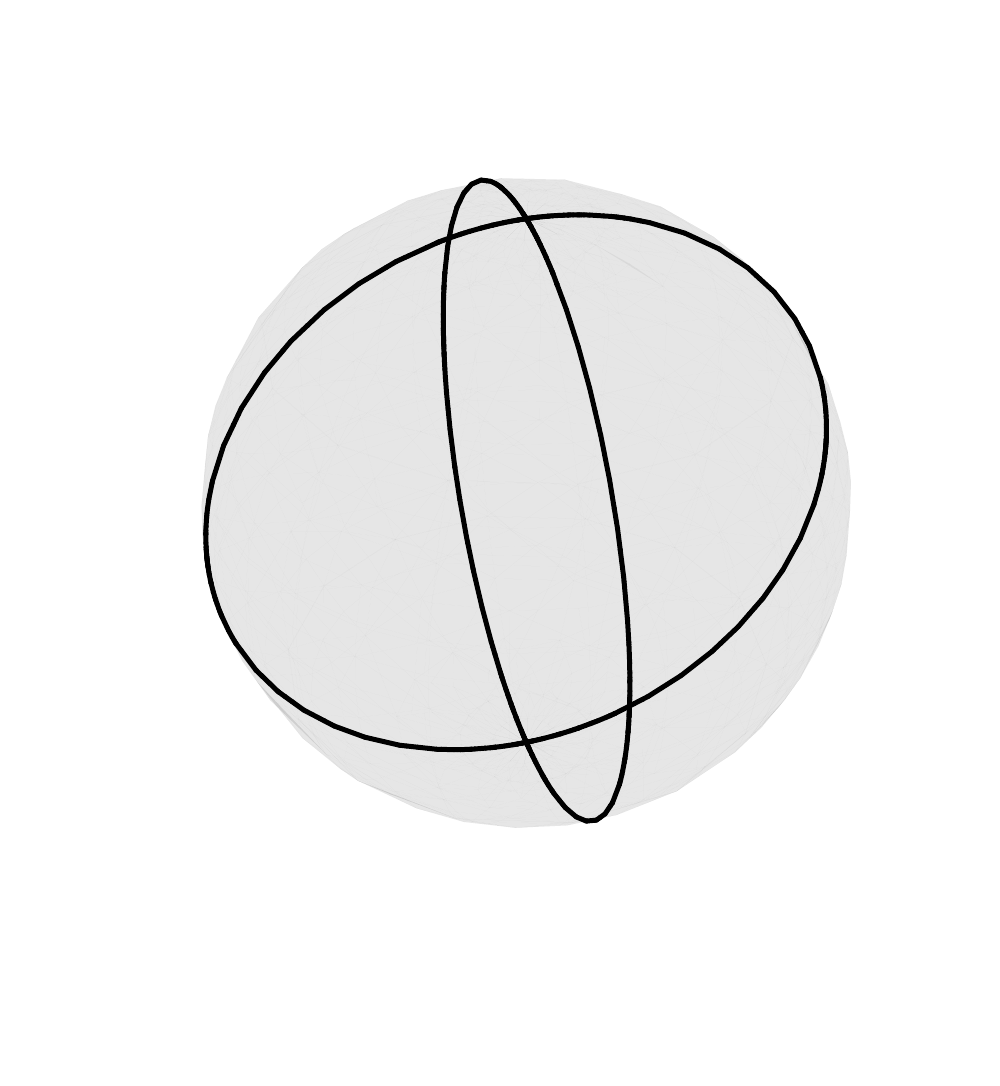}
\caption{Evolution of line nodes for real order parameter $\vec{\Delta}$. The uniaxial case (top left corner) is $\rho=0$, whereas $\rho>0$ corresponds to the biaxial case. The momentum spheres have radius $p_0=(\Delta^2+\mu^2)^{1/4}$ and the black lines indicate the location of the line nodes. We also display the corresponding solutions $\theta_0(\phi)$ to Eq. (\ref{line2}), where the momentum configuration is parametrized as $\textbf{p} =p_0(\cos\phi\sin\theta,\sin\phi\sin\theta,\cos\theta)$. The individual panels correspond to $\rho=0,\ 0.9,\ 1$ (top row, from left to right) and $\rho=1.1,\ 20,\ 500$ (bottom row, from left to right).}
\label{FigNodes}
\end{minipage}
\end{figure}

Assume first that $\phi$ is real. We  can rotate into the eigenframe of the matrix and assume without loss of generality that
\begin{align}
 \label{line1} \phi = \Delta_1M^1+\Delta_2M^2
\end{align}
with real eigenvalues $\Delta_{1,2}$. For momentum configurations $\textbf{p}$ that satisfy $\vec{d}(\textbf{p})\cdot\vec{\Delta}=0$ the excitation spectrum reads $E(\textbf{p})^2=(p^2\pm\sqrt{\Delta^2+\mu^2})^2$ and has nodes at a radial amplitude $p_0=(\Delta^2+\mu^2)^{1/4}$. We thus need to discern whether $\vec{d}\cdot\vec{\Delta}=0$ defines a zero- or one-dimensional set of momenta, corresponding to point or line nodes, respectively. Write $\textbf{p} =p_0(\cos\phi\sin\theta,\sin\phi\sin\theta,\cos\theta)$. Due to the simplification in Eq. (\ref{line1}) we have
\begin{align}
 \nonumber 0 = \vec{d}\cdot\vec{\Delta} &= \Delta_1 \frac{\sqrt{3}}{2}(p_x^2-p_y^2) + \Delta_2 \frac{1}{2}(2p_z^2-p_x^2-p_y^2)\\
 \nonumber &= \Delta_1 \frac{\sqrt{3}}{2}p_0^2 \sin^2\theta\cos(2\phi)+\Delta_2\frac{1}{2}p_0^2(2-3\sin^2\theta)\\
 \label{line2} &\propto \rho \cos(2\phi) + \frac{2}{\sin^2\theta}-3
\end{align}
with $\rho = \sqrt{3}\Delta_1/\Delta_2$ and we assumed $\sin\theta\neq 0$. This equation has parametric solutions $(\theta_0(\phi),\phi)$, where $\theta_0(\phi)$ consists of the two branches
\begin{align}
\label{line3} \theta_0(\phi) &=\begin{cases} \arcsin(\sqrt{\frac{2}{3-\rho\cos(2\phi)}}) \\ \pi - \arcsin(\sqrt{\frac{2}{3-\rho\cos(2\phi)}})\end{cases}.
\end{align}
This corresponds to line nodes for all values of $\rho<\infty$. For $\rho=\infty$ (i.e. $\Delta_2=0$) the solution to Eq. (\ref{line2}) is given by $(\theta,\phi_0(\theta))$ with $\phi_0(\theta)=\frac{\pi}{4},\ \frac{3\pi}{4}$. In Fig. \ref{FigNodes} we display the evolution of the line nodes from the uniaxial ($\rho=0$) to the general biaxial case ($\rho>0$). We observe that the parallel circles of the uniaxial solution start to wiggle for small $\rho$ and eventually become connected for $\rho=1$. For $\rho>1$ the two lines split again and with increasing $\rho$ they deform into the configuration that corresponds to the $\Delta_2=0$ solution of two intersecting circles.

In the case of complex tensor orders satisfying $\vec{\Delta}^2=0$, the momentum configurations with $\vec{d}\cdot\vec{\Delta}=0$ have excitation energies $E(\textbf{p})^2=(p^2\pm\mu)^2$ and $E(\textbf{p})^2=(p^2\pm\sqrt{2\Delta^2+\mu^2})^2$, where the signs are independent. However, the \emph{complex} equation $\vec{d}\cdot\vec{\Delta}=0$ generically only leads to point solutions because it essentially consists of two equations for the two angles $(\theta,\phi)$. These point nodes are inflated, see Ref. \cite{PhysRevLett.118.127001}. A notable exception to point nodes, as pointed out in Ref. \cite{PhysRevLett.116.177001}, is given by $\vec{\Delta}=\frac{\Delta}{\sqrt{2}}(0,0,1,\rmi,0)$, because $\vec{d}\cdot\vec{\Delta} =\sqrt{3}\Delta p_z(p_x+\rmi p_y)=0$ allows for arbitrary $p_{x,y}$ if $p_z=0$, resulting in an equatorial line node.

\section{Ginzburg--Landau free energy}\label{AppGL}
\begin{figure}[t]
\centering
\includegraphics[width=8.5cm]{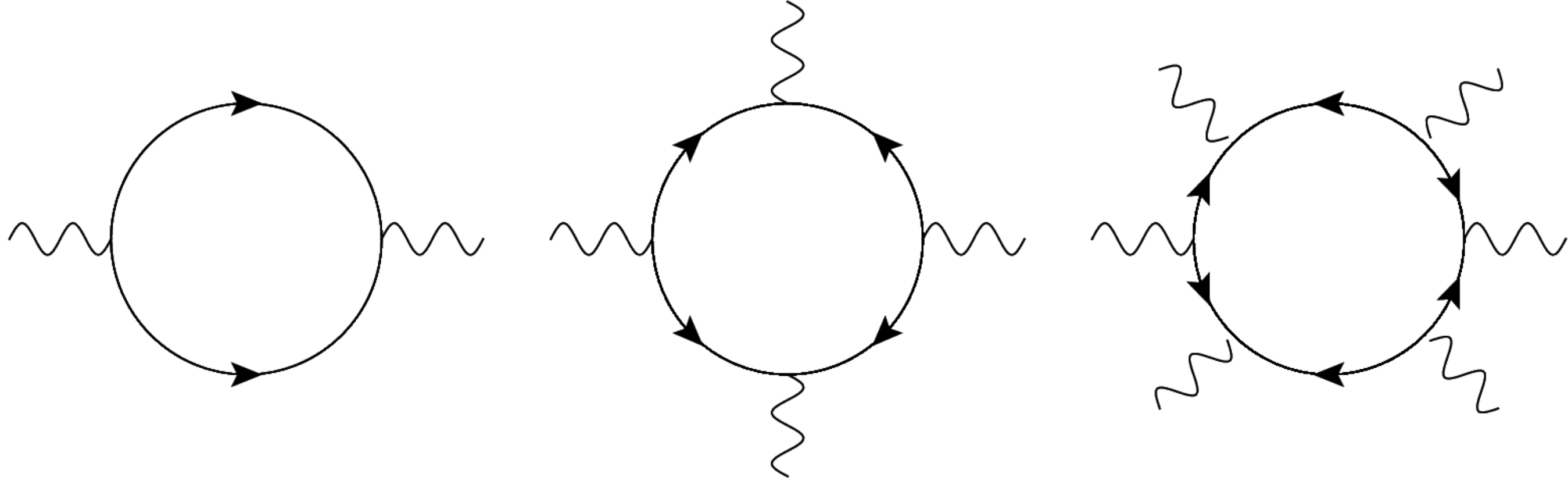}
\caption{Diagrammatic depiction of the loop integration for the functions $K_{ab}$, $K_{abcd}$, $K_{abcdef}$ entering $F_{2,4,6}$ (from left to right). The solid lines constitute fermion propagators, and the wiggly lines are insertions of $\Delta$ or $\Delta^*$ with vertices $\gamma_{45}\gamma_a$, respectively.}
\label{FigLoops}
\end{figure}

We compute the GL free energy $F(\phi)$ as it is employed in the main text. The second-order transition is obtained from a one-loop expansion to sextic order. Let $F_n(\vec{\Delta})$ be the contribution to an expansion of $F(\phi)$ that contains $n$ powers of $\phi$. We have
\begin{align}
 \label{gl1} F_2(\vec{\Delta}) &= \frac{1}{g}|\vec{\Delta}|^2-\frac{1}{2}K_{ab}(\mu,T,\Lambda)\Delta_a^*\Delta_b,\\
 \label{gl2} F_4(\vec{\Delta})  &= \frac{1}{4}K_{abcd}(\mu,T,\Lambda)\ \Delta_a^*\Delta_b\Delta_c^*\Delta_d,\\
 \label{gl3} F_6(\vec{\Delta}) &= -\frac{1}{6}K_{abcdef}(\mu,T,\Lambda)\ \Delta_a^*\Delta_b\Delta_c^*\Delta_d\Delta_e^*\Delta_f,
\end{align}
where the functions $K$ express the loop integrations contained in the diagrams in Fig. \ref{FigLoops}. For their explicit expressions we introduce the propagator
\begin{align}
 \label{gl4} G(Q)= \frac{-\rmi q_0+d_a(\textbf{q})\gamma_a+\mu}{q_0^2+q^4+2\rmi q_0\mu -\mu^2}.
\end{align}
Here $Q=(q_0,\textbf{q})$ with $q=|\textbf{q}|^2$, and $q_0=2\pi(n+1/2)T$ denotes Matsubara frequencies with $n\in\mathbb{Z}$. We write
\begin{align}
 \label{gl5}   \int_{\textbf{q}}^\Lambda (\dots)= \int^\Lambda \frac{\mbox{d}^3q}{(2\pi)^3}(\dots),
\end{align}
where the momentum integration is restricted to the domain $0\leq q \leq \Lambda$. We have
\begin{align}
 \label{gl6} K_{a_1\dots a_{2m}} = T\sum_{n}  \int_{\textbf{q}}^\Lambda \mbox{tr} \Bigl( \prod_{i=1}^m G(Q) \gamma_{a_{2i-1}}G(-Q)\gamma_{a_{2i}}\Bigr).
\end{align}
In writing this expression we employed $\gamma_{45}G(Q)^{\rm T}\gamma_{45}=G(Q)$.

We parametrize the most general terms in $F_n$ by means of
\begin{align}
 \label{gl7} F_2(\vec{\Delta})={}& r |\vec{\Delta}|^2,\\
 \label{gl8} F_4(\vec{\Delta}) ={}& q_1 |\vec{\Delta}|^4+q_2|\vec{\Delta}^2|^2 + q_3\mathcal{Q},\\
 \nonumber F_6(\vec{\Delta}) ={}& s_1 |\vec{\Delta}|^6 + s_2 |\vec{\Delta}|^2|\vec{\Delta}^2|^2 + s_3 |\mbox{tr}(\phi^3)|^2 \\
 \label{gl9} &+ s_4 |\mbox{tr}(\phi^2\phi^\dagger)|^2 + s_5 |\vec{\Delta}|^2\mathcal{Q}
\end{align}
with $\mathcal{Q}=\frac{1}{2}\mbox{tr}(\phi^\dagger\phi\phi^\dagger\phi)$. Choosing the configurations
\begin{align}
 \label{gl10} \vec{\Delta}_1 &= \Delta(1,0,0,0,0),\ \vec{\Delta}_2  = \Delta(0,1,0,0,0),\\
 \label{gl11} \vec{\Delta}_3 &= \frac{\Delta}{\sqrt{2}}(1,\rmi,0,0,0),\ \vec{\Delta}_4 = \frac{\Delta}{\sqrt{2}} (0,0,1,\rmi,0),\\
 \label{gl12} \vec{\Delta}_5 &= \frac{\Delta}{\sqrt{2}}(1,0,\rmi,0,0)
\end{align}
we employ Eqs. (\ref{gl1})-(\ref{gl3}) and $F_2(\vec{\Delta}_1)=r\Delta^2$,
\begin{align}
 F_4(\vec{\Delta}_2) &= (q_1+q_2+q_3)\Delta^4,\\
 F_4(\vec{\Delta}_3) &= \Bigl(q_1+\frac{2}{3}q_3\Bigr)\Delta^4,\\
 F_4(\vec{\Delta}_4) &= (q_1+q_3)\Delta^4,
\end{align}
and
\begin{align}
 F_6(\vec{\Delta}_1) &= (s_1+s_2+s_5)\Delta^6,\\
 F_6(\vec{\Delta}_2) &= \Bigl(s_1+s_2+\frac{4}{3}(s_3+s_4)+s_5\Bigr)\Delta^6,\\
 F_6(\vec{\Delta}_3) &= \Bigl(s_1+\frac{8}{3}s_3+\frac{2}{3}s_5\Bigr)\Delta^6,\\
 F_6(\vec{\Delta}_4) &= (s_1+s_5)\Delta^6,\\
 F_6(\vec{\Delta}_5) &= \Bigl(s_1+\frac{9}{8}s_3+\frac{1}{8}s_4+s_5\Bigr)\Delta^6
\end{align}
to obtain the coefficients $r, q_i, s_i$. The quadratic term $r$ is discussed in the next section, see Eq. (\ref{tc1}). For the quartic and sextic terms we verify
\begin{align}
 \label{gl13} q_3=s_5=0,\ s_4=9s_3.
\end{align}
To display the remaining coefficients we write
\begin{align}
 \label{gl14} q_i(\mu,T,\Lambda) &= \frac{1}{T^{3/2}}f_{q_i}(\mu/T,\Lambda/\sqrt{T}),\\
 \label{gl15} s_i(\mu,T,\Lambda) &= \frac{1}{T^{7/2}}f_{s_i}(\mu/T,\Lambda/\sqrt{T}).
\end{align}
We introduce $\textbf{x}=\textbf{q}/\sqrt{T}$ and $x_0=q_0/T=2\pi(n+1/2)$ and find
\begin{widetext}
\begin{align}
 f_{q_1}(y,\hat{\Lambda}) &=  \int_{\textbf{x}}^{\hat{\Lambda}} \sum_n \frac{2x_0^4+x_0^2( -\frac{36}{5} x^4+4y^2)+\frac{6}{7}x^8-\frac{12}{5}x^4y^2+2y^4}{[x_0^2+(x^2-y)^2]^2[x_0^2+(x^2+y)^2]^2},\\
 f_{q_2}(y,\hat{\Lambda}) &= \int_{\textbf{x}}^{\hat{\Lambda}} \sum_n \frac{-x_0^4+x_0^2(\frac{14}{5}x^4-2y^2)-\frac{27}{35}x^8+2x^4y^2-y^4}{[x_0^2+(x^2-y)^2]^2[x_0^2+(x^2+y)^2]^2},\\
  f_{s_1}(y,\hat{\Lambda})  &=\int_{\textbf{x}}^{\hat{\Lambda}} \sum_n \frac{-\frac{8}{3}x_0^6+x_0^4(24x^4-8y^2)+x_0^2(-\frac{120}{7}x^8+\frac{144}{5}x^4y^2-8y^4)+\frac{13688}{15015}x^{12}-\frac{24}{7}x^8y^2+\frac{24}{5}x^4y^4-\frac{8}{3}y^6}{[x_0^2+(x^2-y)^2]^3[x_0^2+(x^2+y)^2]^3},\\
 f_{s_2}(y,\hat{\Lambda})  &= \int_{\textbf{x}}^{\hat{\Lambda}} \sum_n \frac{2x_0^6+x_0^4(-\frac{82}{5}x^4+6y^2)+x_0^2(\frac{62}{5}x^8-\frac{108}{5}x^4y^2+6y^4)-\frac{862}{1001}x^{12}+\frac{122}{35}x^8y^2-\frac{26}{5}x^4y^4+2y^6}{[x_0^2+(x^2-y)^2]^3[x_0^2+(x^2+y)^2]^3},\\
 f_{s_3}(y,\hat{\Lambda})  &= -\frac{64}{5005} \int_{\textbf{x}}^{\hat{\Lambda}} \sum_n \frac{x^{12}}{[x_0^2+(x^2-y)^2]^3[x_0^2+(x^2+y)^2]^3}.
\end{align}
\end{widetext}
The Matsubara summation over $n$ in each expression can be performed analytically, leaving a one-dimensional integral over $x$. Note that $s_3$ is manifestly negative.

\section{Critical temperature}\label{AppTc}
We derive the analytic expression for the weak coupling critical temperature for nonzero $\mu$ given in the main text. The quadratic term in the expansion of the free energy, $F_2 = r |\vec{\Delta}|^2$, can be written as
\begin{align}
 \label{tc1} r(g,\mu,T,\Lambda) = \frac{1}{g} - \frac{1}{g_{\rm c}} + T^{1/2} f_r(\mu/T,\Lambda/\sqrt{T}),
\end{align}
where $\frac{1}{g_{\rm c}} = \frac{\Lambda}{10\pi^2}$ is the inverse critical coupling, and the function $f_r(y,\hat{\Lambda})$ is given by
\begin{align}
 \nonumber f_r(y,\hat{\Lambda}) &= \int_{\textbf{x}}^{\hat{\Lambda}} \Bigl(\sum_{n} \frac{-2x_0^2+\frac{6}{5}x^4-2y^2}{[x_0^2+(x^2-y)^2][x_0^2+(x^2+y)^2]}\\
 \label{tc2} &+\frac{1}{5x^2}\Bigr).
\end{align}
The notation is adopted from the previous section.

In the following we compute the function $f_r(y,\hat{\Lambda})$ for large $y$. We assume $T>0$ and $y=\frac{\mu}{T}>0$. The integral in Eq. (\ref{tc2}) is ultraviolet finite due to the term $+\frac{1}{5x^2}$ and we can send $\hat{\Lambda}\to \infty$. In order to separate the divergent part of the expression for $y\to\infty$ from the non-divergent one, we decompose the expression according to
\begin{align}
 \nonumber f_r(y) &= \int_{\textbf{x}} \Biggl(\sum_{n} \Bigl(\frac{\frac{6}{5}x^4-2y^2}{4x^2y}+\frac{2(x^2-y)^2}{4x^2y}\Bigr)\frac{1}{x_0^2+(x^2-y)^2} \\
  \nonumber &-\sum_{n}\Bigl(\frac{\frac{6}{5}x^4-2y^2}{4x^2y}+\frac{2(x^2+y)^2}{4x^2y}\Bigr)\frac{1}{x_0^2+(x^2+y)^2}\\
 \label{tc3} &+\frac{1}{5x^2}\Biggr).
\end{align}
Evaluating the Matsubara summation yields
\begin{align}
 \nonumber f_r(y) &= \int_{\textbf{x}}\Bigl[\frac{\frac{4}{5}x^2-y}{y|x^2-y|}\Bigl(\frac{1}{2}-\frac{1}{e^{|x^2-y|}+1}\Bigr)-\frac{2}{5y}+\frac{1}{10x^2}\Bigr]\\
 \nonumber &+\int_{\textbf{x}} \Bigl[-\frac{\frac{4}{5}x^2+y}{y(x^2+y)}\Bigl(\frac{1}{2}-\frac{1}{e^{x^2+y}+1}\Bigr)+\frac{2}{5y}+\frac{1}{10x^2}\Bigr]\\
 \label{tc4} &= f_1(y) + f_2(y) + f_3(y),
\end{align}
with
\begin{align}
 \nonumber f_1(y) &= -\frac{1}{5} \int_{\textbf{x}}\Bigl[ \frac{y}{y|x^2-y|} \Bigl(\frac{1}{2}-\frac{1}{e^{|x^2-y|}+1}\Bigr)-\frac{1}{2x^2}\Bigr],\\
 \label{tc4b}f_2(y) &= \frac{4}{5}  \int_{\textbf{x}} \Bigl[\frac{x^2-y}{y|x^2-y|}\Bigl(\frac{1}{2}-\frac{1}{e^{|x^2-y|}+1}\Bigr)-\frac{1}{2y}\Bigr],\\
  \nonumber f_3(y) &= \int_{\textbf{x}} \Bigl[-\frac{\frac{4}{5}x^2+y}{y(x^2+y)}\Bigl(\frac{1}{2}-\frac{1}{e^{x^2+y}+1}\Bigr)+\frac{2}{5y}+\frac{1}{10x^2}\Bigr].
\end{align}
The three individual terms are ultraviolet finite and can be evaluated separately. For the first term we employ
\begin{align}
 \label{tc5} \int_0^\infty \mbox{d}z \Bigl[\frac{z^2}{|z^2-1|}\tanh\Bigl(\frac{y|z^2-1|}{2}\Bigr)-1\Bigr] \to \log\Bigl(\frac{8e^{\gamma-2}}{\pi}y\Bigr)
\end{align}
for $y\to\infty$, as is well-known from textbook BCS theory, and, consequently,
\begin{align}
 \nonumber f_1(y)&= -\frac{\sqrt{y}}{20\pi^2} \int_0^\infty \mbox{d}z \Bigl[\frac{z^2}{|z^2-1|}\tanh\Bigl(\frac{y|z^2-1|}{2}\Bigr)-1\Bigr]\\
 \label{tc6} &\to -\frac{\sqrt{y}}{20\pi^2}\log\Bigl(\frac{8e^{\gamma-2}}{\pi}y\Bigr).
\end{align}
For the second contribution we employ
\begin{align}
\int_0^\infty \mbox{d}z\frac{-z^2}{e^{y(z^2-1)}+1} &= \frac{\sqrt{\pi}}{4y^{3/2}} \text{Li}_{3/2}(-e^y)\\
 &\to \frac{\sqrt{\pi}}{4y^{3/2}} \frac{-y^{3/2}}{\Gamma(5/2)}=-\frac{1}{3},
\end{align}
where $\text{Li}_\nu(z)= \sum_{k=1}^\infty\frac{z^k}{k^\nu}$ is the polylogarithm, and we used $\text{Li}_\nu(-e^y)\to -y^\nu/\Gamma(\nu+1)$ for $y\to\infty$ and $\nu>0$. Hence,
\begin{align}
 \nonumber f_2(y) &= \frac{4}{5} \int_{\textbf{x}} \Bigl[\frac{x^2-y}{y|x^2-y|}\Bigl(\frac{1}{2}-\frac{1}{e^{|x^2-y|}+1}\Bigr)-\frac{1}{2y}\Bigr]\\
 \nonumber &= \frac{4}{5} \int_{\textbf{x}} \Bigl[\frac{x^2-y}{y(x^2-y)}\Bigl(\frac{1}{2}-\frac{1}{e^{x^2-y}+1}\Bigr)-\frac{1}{2y}\Bigr]\\
 \nonumber &=\frac{4}{5y} \int_{\textbf{x}} \Bigl[\frac{1}{2}-\frac{1}{e^{x^2-y}+1}-\frac{1}{2}\Bigr]\\
 \label{tc7}  &= \frac{4}{5} \frac{\sqrt{y}}{2\pi^2}\int_0^\infty \mbox{d}z\frac{-z^2}{e^{y(z^2-1)}+1}\to-\frac{2\sqrt{y}}{15\pi^2}.
\end{align}
At last, the third contribution is convergent for $T\to 0$, and we can boldly send $y\to \infty$ to arrive at
\begin{align}
 \nonumber f_3(y) &= \frac{\sqrt{y}}{2\pi^2} \int_0^\infty\mbox{d}z\ z^2 \Bigl[-\frac{\frac{4}{5}z^2+1}{z^2+1}\Bigl(\frac{1}{2}-\frac{1}{e^{y(z^2+1)}+1}\Bigr)\\
 \nonumber &+\frac{2}{5}+\frac{1}{10z^2}\Bigr]\\
 \nonumber &\to  \frac{\sqrt{y}}{2\pi^2} \int_0^\infty\mbox{d}z\ z^2 \Bigl[-\frac{\frac{4}{5}z^2+1}{2(z^2+1)}+\frac{2}{5}+\frac{1}{10z^2}\Bigr]\\
 \label{tc8} &= \frac{\sqrt{y}}{2\pi^2} \frac{\pi}{20} = \frac{\sqrt{y}}{40\pi}.
\end{align}
Taken together we arrive at
\begin{align}
 \label{tc9} f_r(y) \to -\frac{\sqrt{y}}{20\pi^2} \log\Bigl(\frac{8e^{\gamma+\frac{2}{3}-\frac{\pi}{2}}}{\pi}y\Bigr)
\end{align}
for $y\to\infty$.

The weak coupling gap equation for $T/\mu\to 0$ and $\Lambda^2/T\to\infty$ becomes
\begin{align}
 \label{tc10} 0 = \frac{1}{g}-\frac{1}{g_{\rm c}} -\frac{\sqrt{\mu}}{20\pi^2}  \log\Bigl(\frac{8e^{\gamma+\frac{2}{3}-\frac{\pi}{2}}}{\pi}\frac{\mu}{T_{\rm c}}\Bigr),
\end{align}
which is solved by
\begin{align}
 \nonumber \frac{T_{\rm c}}{\mu} &= \frac{8e^{\gamma+\frac{2}{3}-\frac{\pi}{2}}}{\pi} \exp\Bigl\{-\frac{1-g/g_{\rm c}}{g/g_{\rm c}} \frac{20\pi^2}{\sqrt{\mu}g_{\rm c}}\Bigr\}\\
 \label{tc11} &= \frac{8e^{\gamma+\frac{2}{3}-\frac{\pi}{2}}}{\pi} \exp\Bigl\{-\frac{1-g/g_{\rm c}}{g/g_{\rm c}} \frac{2\Lambda}{\sqrt{\mu}}\Bigr\}.
\end{align}
This proves the formula given in the main text. The numerical value of the prefactor of the exponential is $1.836$.

\end{document}